\documentclass[aps,pra,onecolumn,superscriptaddress,floatfix]{revtex4-2}
\usepackage[dvips]{color}
\usepackage{graphicx}
\usepackage{amssymb}
\usepackage{amsmath}
\usepackage{amsfonts}
\usepackage{bbold}

\usepackage{soul}
\usepackage{xcolor}

\begin{document}

\title{Symmetries of the squeeze-driven Kerr oscillator}

\author{Francesco Iachello}

\affiliation{Center for Theoretical Physics, Sloane Physics Laboratory,
Yale University, New Haven, CT 06520-8120, USA}

\author{Rodrigo G. Corti\~nas}
\affiliation{Yale Quantum Institute, Department of Applied Physics and Physics, Yale University, New Haven, Connecticut 06520, USA}

\author{Francisco P\'erez-Bernal}
\affiliation{Depto. de Ciencias Integradas y Centro de Estudios Avanzados en F\'isica, Matem\'aticas y Computaci\'on, Universidad de Huelva, Huelva 21071, Spain}
\affiliation{Instituto Carlos I de F\'isica Te\'orica y Computacional, Universidad de Granada, Fuentenueva s/n, 18071 Granada, Spain}

\author{Lea F. Santos}
\affiliation{Department of Physics, University of Connecticut, Storrs, Connecticut 06269, USA}

\begin{abstract}
We study the symmetries of the static effective Hamiltonian of a driven superconducting nonlinear oscillator, the so-called squeeze-driven Kerr Hamiltonian, and discover a remarkable quasi-spin symmetry $su(2)$ at integer values of the ratio $\eta=\Delta /K$ of the detuning parameter $\Delta $ to the Kerr coefficient $K$. We investigate the stability of this newly discovered symmetry to high-order perturbations arising from the static effective expansion of the driven Hamiltonian. Our finding may find applications in the generation and stabilization of states useful for quantum computing. Finally, we discuss other Hamiltonians with similar properties and within reach of current technologies.
\end{abstract}

%


\maketitle
%
%

\section{Introduction}
\label{sec_intro}

Kerr-nonlinear parametric oscillators (KPOs) have been suggested as devices for quantum computation~\cite{Goto2016,goto2019quantum}. A KPO can generate Schr\"odinger cat states via quantum adiabatic evolution through its bifurcation point. These states correspond to quantum superpositions of coherent states and they are protected against photon dephasing errors, which has motivated their use as the logical states of a qubit~\cite{mirrahimi2014}.
A considerable amount of effort has gone in the last few years in developing KPOs that can generate cat states deterministically, especially with superconducting circuits with Josephson junctions ~\cite{puri2017,grimm2020,Blais2021,darmawan_practical_2021,Kwon2022,Frattini2022,Venkatraman2022_Exp,VenkatramanThesis}. 
In these circuits, the non-linearity of the Josephson junctions is used for achieving large Kerr effects~\cite{Kirchmair2013,Rehak2014,puri2017quantum,Goto2018, Amin2018} and the magnetic flux of a superconducting quantum interference device
(SQUID) is used for the parametric modulation~\cite{Yamamoto2008,Bourassa2012,Wustmann2013,Krantz2013,Eichler2014,Lin2014,Krantz2016}.

Theoretical studies~\cite{Dykman1990,Marthaler2007,DykmanBook2012,Peano2012,Lin2015,Goto2016b,Zhang2017,dykman2018interaction,Roberts2020}  have inspired the analysis of the results of several experimental implementations of single KPOs~\cite{Kwon2022,Frattini2022,Venkatraman2022_Exp}.
A useful tool for these studies is the conversion of  the time-dependent Hamiltonian describing the experimental system into a static effective Hamiltonian. Recently, a general method for this conversion has been developed~\cite{Frattini2022,Venkatraman2022_PRL,xiao2023diagrammatic,VenkatramanThesis},  wherein the static effective Hamiltonian is obtained as a boson expansion in terms of one-dimensional boson annihilation and creation operators, $\hat{a}$ and $\hat{a}^{\dagger}$, with $[\hat{a},\hat{a}^{\dagger}]=1$. To second order in the boson expansion, the static effective Hamiltonian of a driven superconducting nonlinear (Kerr) oscillator in the quantum regime was obtained as~\cite{Frattini2022,Venkatraman2022_Exp},
\begin{equation}
\hat{H} = -\Delta \hat{a}^\dagger \hat{a}  + K\hat{a}^{\dagger 2} \hat{a}^2 - \varepsilon_2(\hat{a}^{\dagger 2}  + \hat{a}^2)~,
\label{effective_H}
\end{equation}
where the detuning $\Delta$, the Kerr coefficient $K$, and the squeezing amplitude $\varepsilon_2$ are explicit functions of the parameters of a driven quantum circuit. We note that, for convenience, Hamiltonian  (\ref{effective_H}) differs from that in \cite{Frattini2022,Venkatraman2022_Exp} by an overall minus sign. This change in sign was also made in Ref.~\cite{goto2019quantum} to conform with standard quantum computing notation.

The spectrum of the driven system was experimentally measured as a function of the control parameters $\Delta $, $K$, and $\varepsilon_{2}$ \cite{Frattini2022} and was found to be accurately described by the second-order Hamiltonian (\ref{effective_H}), although deviations may occur for large nonlinearities~\cite{nacho}.
The observed spectrum shows remarkable properties for integer values of the ratio $\eta = \Delta/K$, which  persist even when the squeezing amplitude is increased to non-perturbative values~\cite{Venkatraman2022_Exp}. Specifically, the spectrum presents real crossings when $\eta$ is even and avoided crossings when $\eta$ is odd \cite{Venkatraman2022_Exp}, which implies that by tuning the  parameters of the system, one can suppress or enhance quantum tunneling~\cite{Venkatraman2022_Exp,Prado2023}.
The spectrum also exhibits an excited state quantum phase transition (ESQPT) as a function of the squeezing amplitude $\varepsilon_{2}$~\cite{Prado2023,Chavez2022}. This ESQPT is similar to the QPTs observed in other systems, such as driven Rabi and Dicke models~\cite{Shen2022,Yang2023} and the Jaynes-Cummings model~\cite{Shen2022}.

In this article, we uncover the symmetries of the squeeze-driven Kerr
oscillator, in particular, the one that occurs as a function of $\Delta$ and which may play an important role in the generation of stable states for quantum
computing \cite{Venkatraman2022_Exp,ruiz2022}. The symmetry, which occurs for integer values of the dimensionless
parameter $\eta =\Delta /K$, is a dynamic symmetry \cite{Iachello1994} similar to
those observed in the interacting boson model of nuclear physics \cite{booknuc}
and the vibron model of molecular physics \cite{bookmol}. Dynamic symmetries
are situations in which the eigenvalues of $\hat{H}$ can be written in terms
of quantum numbers labelling the irreducible representations of an algebra $g\supset g^{\prime }\supset ...$ and often display degeneracies associated
with the subalgebras $g^{\prime }\supset ...$ of $g$ \cite{bookalg}. They
have played a major role in the analysis of data in a variety of fields,
including molecular, atomic, nuclear, and particle physics. The symmetry
which occurs for integer values of $\eta=\Delta /K$ can also be labelled a
``local'' symmetry, since it occurs for special values in the parameter space.
It differs from the ``global'' symmetry parity, $\Pi \equiv Z_{2}$, of the
Hamiltonian of the squeeze-driven Kerr oscillator (\ref{effective_H}), which occurs for any
value in the parameter space. (The global $Z_{2}$ symmetry of the Kerr oscillator (\ref{effective_H}) and of similar models has been investigated by many authors, especially in connection with Limbladian operator obtained from the Hamiltonian operator~\cite{Braak2011,Albert2014}.)

After the identification in sections~\ref{sec_algstruct} and \ref{sec_altalgstruct} of the spectrum generating
algebra of the problem as the symplectic algebra $sp(2,\mathbb{R})$ and of its
alternative Heisenberg algebra $h(2)$, we introduce, in subsection~\ref{subsect_symm_1}, the novel
and unexpected quasi-spin symmetry $su(2)$ of the Kerr oscillator $\hat{H} = -\Delta \hat{a}^{\dag }\hat{a}+K\hat{a}^{\dag ^{2}}\hat{a}^{2}$. We then
return to the squeeze-driven Kerr oscillator of (\ref{effective_H}) and discuss, in subsections~\ref{subsect_symm_2} and \ref{subsect_symm_3}, its
relationship with the theory of ESQPTs \cite{Caprio2008, Cejnar2008, Cejnar2021}. In section~\ref{sect_stability}, we discuss other Hamiltonians with similar properties and within the reach of current technologies. Conclusions are presented in section~6.

\section{Algebraic structure of the squeeze-driven Kerr oscillator}
\label{sec_algstruct}

To reveal the symmetries of the squeeze-driven Kerr Hamiltonian~\cite{Frattini2022,Venkatraman2022_Exp}, we rewrite Hamiltonian (\ref{effective_H}) as 
\begin{equation}
\hat{H}=-\Delta (\hat{a}^{\dag }\hat{a})+K(\hat{a}^{\dag }\hat{a})(\hat{a}^{\dag }\hat{a}-1)-\varepsilon _{2}(\hat{a}^{\dag}\hat{a}^{\dag }+\hat{a}\hat{a})~,
\label{Eq:a+aa+a}
\end{equation}
or, in short,
\begin{equation}
\hat{H}=-\Delta \hat{n}+K\hat{n}(\hat{n}-1)-\varepsilon_{2}\hat{P}_{2}~,
\end{equation}
where $\hat{n}=\hat{a}^{\dag }\hat{a}$ is the number operator and $\hat{P}
_{2}=\hat{a}^{\dag }\hat{a}^{\dag }+\hat{a}\hat{a}$ is the pairing operator
of order two. 

The three operators in (\ref{Eq:a+aa+a}), 
\begin{equation}
\hat{F}_{+}=\hat{a}^{\dag }\hat{a}^{\dag }~,~~\hat{F}_{-}=\hat{a}\hat{a}~, ~~ \hat{F}_{z}=\hat{a}^{\dag }\hat{a} ,
\end{equation}
form a closed algebra with commutation relations
\begin{equation}
\left[ \hat{F}_{z},\hat{F}_{\pm }\right] =\pm 2\hat{F}_{\pm }~,~~ 
\left[ \hat{F}_{+},\hat{F}_{-}\right] =-4\hat{F}_{z}-2~.
\end{equation}
This algebra is the symplectic algebra $sp(2,\mathbb{R})$ \cite{Gilmore1974}. The algebra $sp(2,\mathbb{R})$ is
isomorphic to $su(1,1)$, the non-compact version of $su(2)$ \cite{bookalg},
as one can see by considering the operators 
\begin{equation}
\hat{F}_{+}^{^{\prime \prime }}=\frac{1}{2}(\hat{a}^{\dag }\hat{a}^{\dag })~,~~ \hat{F}_{-}^{^{\prime \prime }}=\frac{1}{2}(\hat{a}\hat{a})~,~~\hat{F}_{z}^{^{\prime \prime }}=\frac{1}{2}\left(\hat{a}^{\dag }\hat{a}+\frac{1}{2}\right)~,
\end{equation}
satisfying the standard form of the commutation relations of $su(1,1)$
\begin{equation}
\left[ \hat{F}_{z}^{^{\prime \prime }}, \hat{F}_{\pm }^{^{\prime \prime }}
\right] =\pm \hat{F}_{\pm }^{^{\prime \prime }}~,~~ \left[ \hat{F}
_{+}^{^{\prime \prime }},\hat{F}_{-}^{^{\prime \prime }}\right] =-2\hat{F}
_{z}^{^{\prime \prime }}~.
\end{equation}
Since the Hamiltonian is written in terms of elements of this algebra, 
\begin{equation}
\hat{H}=-\Delta \hat{F}_{z}+K\hat{F}_{z}(\hat{F}_{z}-1)-\varepsilon
_{2}\left( \hat{F}_{+}+\hat{F}_{-}\right)~,
\end{equation}%
$sp(2,\mathbb{R})$ is the spectrum generating algebra \cite{Iachello1994} of the
Kerr problem at first and second orders. Also, introducing $\hat{F}_{\pm }=\hat{F}_{x}\pm i\hat{F}_{y}$, the last term can be written as $\hat{F}_{+}+\hat{F}_{-}=2\hat{F}_{x}$.

The boson expansion of the effective Hamiltonian of a squeeze-driven Kerr oscillator was carried out to orders three and four in~\cite{Frattini2022}, since these
terms can also be experimentally implemented with the appropriate choice of the parameters in the time-dependent Hamiltonian.
At third order, the additional contributions to the effective Hamiltonian  can be
written as \cite{Frattini2022, Venkatraman2022_Exp}
\begin{equation}
\hat{H}^{(3)}\!=\!-\Delta ^{(3)}(\hat{a}^{\dag }\hat{a})-\!K^{(3)}(\hat{a}^{\dag }
\hat{a})(\hat{a}^{\dag }\hat{a}-1)+\varepsilon _{2}^{(3)}(\hat{a}^{\dag 2}+
\hat{a}^{2}) 
+\varepsilon _{2}^{\prime }\left(\hat{a}^{\dag 2}(\hat{a}^{\dag }\hat{
a}) 
+(\hat{a}^{\dag }\hat{a})\hat{a}^{2}\right),
\label{order_3}
\end{equation}
and at fourth order as
\begin{equation}
\hat{H}^{(4)} \!=\!-\Delta ^{(4)}(\hat{a}^{\dag }\hat{a})-K^{(4)}(\hat{a}
^{\dag }\hat{a})(\hat{a}^{\dag }\hat{a}-1)  
-\lambda ^{(4)}\!\left((\hat{a}^{\dag }\hat{a})^{3}\!-3(\hat{a}^{\dag }\hat{a}
)^{2}\!-2(\hat{a}^{\dag }\hat{a})\right)+\varepsilon _{4}^{(4)}(\hat{a}^{\dagger 4}+\hat{a}
^{4}).
\label{order_4}
\end{equation}
The last term in (\ref{order_4}) can also be rewritten as 
\begin{equation}
\hat{a}^{\dag 4}+\hat{a}^{4}=(\hat{a}^{\dag 2}+\hat{a}^{2})^{2}-2(\hat{a}
^{\dag }\hat{a})(\hat{a}^{\dag }\hat{a}+1)-2.
\end{equation}
Since again the Hamiltonian contributions $\hat{H}^{(3)}$ and $\hat{H}^{(4)}$
are written in terms of elements of $sp(2,\mathbb{R})$, 
\begin{eqnarray}
\hat{H}^{(3)}&=&-\Delta ^{(3)}\hat{F}_{z}-K^{(3)}\hat{F}_{z}(\hat{F}
_{z}-1)+\varepsilon _{2}^{(3)}(\hat{F}_{+}+\hat{F}_{-})+\varepsilon
_{2}^{\prime }(\hat{F}_{+}\hat{F}_{z}+\hat{F}_{z}\hat{F}_{-})~,\\
\hat{H}^{(4)}&=&-\Delta ^{(4)}\hat{F}_{z}-K^{(4)}\hat{F}_{z}(\hat{F}
_{z}-1)-\lambda ^{(4)}(\hat{F}_{z}^{3}-3\hat{F}_{z}^{2}-2\hat{F}
_{z})+\varepsilon _{4}^{(4)}(\hat{F}_{+}^{2}+\hat{F}_{-}^{2}),   
\end{eqnarray}
this algebra is the spectrum generating algebra of the Kerr oscillator at order four, that is the Hamiltonian $\hat{H} = \sum_{1\leq n\leq 4} \hat{H}^{(n)}$, where $n$ denotes the order in the perturbation parameter \cite{Frattini2022,Venkatraman2022_PRL}, is a polynomial in the elements $\hat{a}^{\dag }\hat{a},\hat{a}^{\dag }\hat{a}^{\dag },\hat{a}\hat{a}$ of the $sp(2,\mathbb{R})$ Lie algebra.

\section{Alternative algebraic structure}
\label{sec_altalgstruct}

An alternative spectrum generating algebra is obtained by introducing an auxiliary boson $s$ \cite{bookalg} and constructing the algebra of $u(2)$ as 
\begin{equation}
\hat{F}_{-}^{\prime }=\hat{s}^{\dag }\hat{a}~,~~ \hat{F}_{+}^{\prime }=\hat{a}^{\dag }\hat{s}~,~~ \hat{F}_{z}^{\prime }=\frac{1}{2}(\hat{s}^{\dag }\hat{s}-\hat{a}^{\dag }\hat{a})~,~~\hat{N}=\hat{s}^{\dag }\hat{s}+\hat{a}^{\dag }\hat{a}~~.
\end{equation}
The three operators $\hat{F}_{+}^{\prime },\hat{F}_{-}^{\prime },\hat{F}
_{z}^{\prime }$ satisfy the commutation relations of the Lie algebra $su(2)$,
\begin{equation}
\left[ \hat{F}_{z}^{\prime }, \hat{F}_{\pm }^{\prime }\right] =\pm \hat{F}
_{\pm }^{\prime }~,~~ \left[ \hat{F}_{+}^{\prime },\hat{F}_{-}^{\prime
}\right] =2\hat{F}_{z}^{\prime }~~.
\end{equation}
Together with $\hat{N}$, they are the elements of the Lie algebra of $u(2)$.

We introduce now the operators $\hat{F}_{-}^{\prime }=\hat{s}^{\dag }\hat{a}$, $
\hat{F}_{+}^{\prime }=\hat{a}^{\dag }\hat{s}$, $\hat{n}=\hat{a}^{\dag }\hat{a
}$, $\hat{n}_{s}=\hat{s}^{\dag }\hat{s}$, replace the operators $\hat{s}$
and $\hat{s}^{\dag }$ by $\sqrt{N}$, and consider the operators 
\begin{equation}
\frac{\hat{
F}_{-}^{\prime }}{\sqrt{N}}=\hat{a}~,~~ \frac{\hat{F}_{+}^{\prime }}{\sqrt{N}}
=\hat{a}^{\dag }~,~~ \hat{n}=\hat{a}^{\dag }\hat{a}~,~~ \frac{\hat{n}_{s}}{N}
=\hat{I}.
\end{equation}
The operators $\hat{a}$,  $\hat{a}^{\dag }$,  $\hat{a}^{\dag }\hat{a}$, and the identity operator, $\hat{I}$, form
an algebra called the Heisenberg algebra, $h(2)$, with commutation relations
\begin{eqnarray}
\left[ \hat{a},\hat{a}^{\dag }\right] &=&\hat{I}, \hspace{0.4 cm} \left[ \hat{a}^{\dag },\hat{I}
\right] =\left[ \hat{a},\hat{I}\right] =0~,  \nonumber \\
\left[ \hat{a},\hat{a}^{\dag }\hat{a}\right] &=&\hat{a}~~, \hspace{0.4 cm}\left[ \hat{a
}^{\dag },\hat{a}^{\dag }\hat{a}\right] =-\hat{a}^{\dag }~.
\end{eqnarray}
The algebra $h(2)$ is called the contracted algebra of $u(2)$ \cite{bookalg},
\begin{equation}
u(2)\rightarrow _{c}h(2)~.
\end{equation}
The algebra $h(2)$ is an alternative spectrum generating algebra  of
the squeeze-driven Kerr oscillator. Calculations for the eigenvalues and
eigenvectors of the squeeze-driven Kerr oscillator can therefore also be done
making use of the algebra $u(2)$ in the limit $N\rightarrow \infty $. The
Hamiltonian $\hat{H}$ at orders 1 and 2 can be rewritten in the $u(2)$ basis
as
\begin{equation}
\hat{H}=-\Delta \hat{n}+K\hat{n}(\hat{n}-1)-\varepsilon _{2}\left( \hat{a}
^{\dag }\hat{a}^{\dag }\hat{s}\hat{s}+\hat{s}^{\dag }\hat{s}^{\dag }\hat{a}
\hat{a}\right)~,
\label{u2_Hamiltonian}
\end{equation}
with contracted form 
\begin{equation}
\hat{H}=-\Delta \hat{n}+K\hat{n}(\hat{n}-1)-\varepsilon _{2}^{\prime }(\hat{a
}^{\dag }\hat{a}^{\dag }+\hat{a}\hat{a})~,
\end{equation}
where $\varepsilon _{2}^{\prime }=\varepsilon _{2}N$. Hamiltonians as in (\ref{u2_Hamiltonian}) were considered years ago \cite{onophd} and are used
in the algebraic approach to stretching vibrations of molecules \cite{bookmol}.

\section{Symmetry and classification of states}
\label{sec_symmetry}

The Hamiltonian (\ref{effective_H}) has a remarkable set of symmetries. For purposes
of studying these symmetries, it is convenient to divide by a scale $K$ and
consider the dimensionless Hamiltonian
\begin{equation}
\frac{\hat{H}}{K}=-\eta \hat{n}+\hat{n}(\hat{n}-1)-\xi \hat{P}_{2}=-\eta ^{\prime }
\hat{n}+\hat{n}^{2}-\xi \hat{P}_{2}~,
\label{Kerr_Squeezed}
\end{equation}
where $\eta =\Delta /K$ and $\xi =\varepsilon _{2}/K$ are control parameters
and $\eta ^{\prime }=\eta +1$. In what follows, we analyze first the symmetries of parts of the Hamiltonian (\ref{Kerr_Squeezed}), namely $-\protect\eta\hat{n}+\hat{n}(\hat{n}-1)$ in subsection 4.1 and $\hat{n}(\hat{n}-1)-\protect\xi \hat{P}_{2}$ in subsection 4.2, before investigating the complete Hamiltonian in subsection 4.3.

\subsection{Symmetries of the Hamiltonian $-\protect\eta\hat{n}+\hat{n}(\hat{n}-1)$}
\label{subsect_symm_1}

\begin{figure}[h]
    \centering
    \includegraphics[scale=0.8]{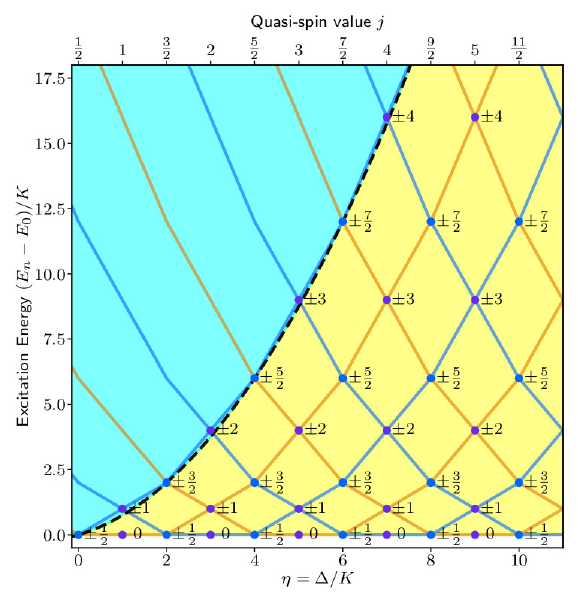}
    \caption{Excitation energy of the Hamiltonian (\ref{Kerr_no_squeezing}) as a function of the control parameter $\eta = \Delta/K$ (ratio of detuning to Kerr coefficient). The quasi-spin $| j,m \rangle $ classification of the degenerate states  with energies $E(\eta =\textrm{odd})=m^{2}$, $E(\eta =\textrm{even})=m^{2}-\frac{1}{4}$ is included. Crossings are marked with blue dots and labelled by their associated $m$ values. The separatrix, marked with a dashed black line, is $E_{s}=\eta/2+\eta ^{2}/4$. The parity of the eigenstates is positive (orange) and negative (blue). The two phases are marked by blue and yellow backgrounds.}
    \label{fig_1}
\end{figure}

We consider first the Hamiltonian 
\begin{equation}
\hat{H}_{1}=\frac{\hat{H}}{K}=-(\eta +1)\hat{n}+\hat{n}^{2}=-\eta ^{\prime }\hat{n}+\hat{n}^{2}~.
\label{Kerr_no_squeezing}
\end{equation}
The spectrum of eigenvalues of this Hamiltonian, counted from the lowest state, is shown in figure~\ref{fig_1}. It is divided into two parts (phases) with separatrix $E_{s}=\eta /2+\eta^{2}/4$ marked in the figure with a dashed black line.

To the left of the separatrix (blue filled region), states are singly degenerate with $
u(1)$ symmetry, $u(1)\doteqdot \hat{a}^{\dag }\hat{a}$, and eigenvalues
\begin{equation}
E=-(\eta +1)n+n^{2}=-\eta ^{\prime }n+n^{2}~~.
\end{equation}
To the right of the separatrix (yellow filled region) and for $\eta ^{\prime }=\eta +1 = \textrm{integer}$,
degeneracies occur. These degeneracies are due to a remarkable (and hitherto unknown) quasi-spin symmetry $su(2)$. 

The degeneracy points can be
characterized by quasi-spin quantum numbers $| j,m \rangle $.
The values of the quasi-spin are $j=\frac{\eta ^{\prime }}{2}=\frac{\eta +1}{2}$, as given in table~\ref{tab_1}.
For each $j$ value, the values of $m$ are 
\begin{eqnarray}
m &=&\pm j,\pm (j-1),...,\pm 1/2~;~~  j=\textrm{half-integer}~;~~  \eta
=\textrm{even}~,  \nonumber \\
m &=&\pm j,\pm (j-1),...,0~;~~  j=\textrm{integer}~;~~  \eta =\textrm{odd}~.
\end{eqnarray}
The eigenvalues $E$ of $\hat{H}/K$, counted from the lowest state, are
\begin{equation}
E =m^{2}-\frac{1}{4}~;~~m = \pm \frac{1}{2},\pm \frac{3}{2},\pm 
\frac{5}{2},...,\pm j~~,
\label{spectrum_nodelta_1}
\end{equation}
for half integer $j$ (even $\eta$) and 
\begin{equation}
E =m^{2}~;~~m=0,\pm 1,\pm 2,...,\pm j~~,
\label{spectrum_nodelta_2}
    \end{equation}
for  integer $j$ (odd $\eta$). Both sets of eigenvalues correspond to the dynamic symmetry $su(2)\supset so(2)$. Note that each
eigenvalue is doubly degenerate, $\pm m$, except for $m=0$, when it is singly
degenerate. This result can be verified from figure \ref{fig_1} where the values of $j$
and $m$ at the degeneracy points are shown.

\begin{table}[h]
\begin{center}
\begin{tabular}{ c|cccccc }
$\eta ^{\prime }$ & 1 & 2 & 3 & 4 & 5 & ... \\ 
$\eta$ & 0 & 1 & 2 & 3 & 4 & ... \\ 
$j$ & 1/2 & 1 & 3/2 & 2 & 5/2 & ...
\end{tabular}
\caption{The values of the quasi-spin are $j=\frac{\eta ^{\prime }}{2}=\frac{\eta +1}{2}$. }
\label{tab_1}
\end{center}
\end{table}

To elucidate this quasi-spin symmetry, it is convenient to
construct the representations $| j,m \rangle $ with two boson
operators $\hat{b}_{1}, \hat{b}_{2}$ (see \cite{bookalg}) with eigenvalues of the
number operators $\hat{n}_{1},\hat{n}_{2}$ satisfying $n_{1}+n_{2}=N$. The
values of $j$ and $m$ are
\begin{equation}
j=\left( \frac{n_{2}+n_{1}}{2}\right), \hspace{0.4 cm} m=\left( \frac{n_{2}-n_{1}}{2}\right) .
\label{eq:jm}
\end{equation}
For example, the values of $m$ for the representation $n_{2}+n_{1}=N=5$, $
j=5/2$, $\eta ^{\prime }=5$, $\eta =\mbox{even}=4$ are given in table \ref{tab_2}.
\begin{table}[h]
\begin{center}
\begin{tabular}{ccccc}
$n_{1}$ & $n_{2}$ & $m$ & $m^{2}$ & $m^{2}-1/4$ \\ 
\hline  
0 & 5 & 5/2 & 25/4 & 6 \\ 
1 & 4 & 3/2 & 9/4 & 2 \\ 
2 & 3 & 1/2 & 1/4 & 0 \\ 
3 & 2 & -1/2 & 1/4 & 0 \\ 
4 & 1 & -3/2 & 9/4 & 2 \\ 
5 & 0 & -5/2 & 25/4 & 6%
\end{tabular}
\caption{Values of $m$ for the representation $n_{2}+n_{1}=N=5$, $
j=5/2$, $\eta ^{\prime }=5$, $\eta =\mbox{even}=4$.}\label{tab_2}
\end{center}
\end{table}
One can verify from figure~\ref{fig_1} that the doubly degenerate states at $\eta^{\prime }=5,\eta =\mbox{even}=4$ have precisely the values of $n_{1},n_{2}$ as
given in table~\ref{tab_2}. 
Similarly, the values of $m$ for the representation $n_{2}+n_{1}=N=4$, $\eta ^{\prime }=4$, $\eta =\mbox{odd}=3$ are given in table~\ref{tab_3} and are precisely those in  figure~\ref{fig_1}, with a singly degenerate state at zero
energy and doubly degenerate states with energy given by $m^{2}$. 
\begin{table}[h]
\begin{center}
    \begin{tabular}{cccc}
$n_{1}$ & $n_{2}$ & $m$ & $m^{2}$ \\
\hline 
0 & 4 & 2 & 4 \\ 
1 & 3 & 1 & 1 \\ 
2 & 2 & 0 & 0 \\ 
3 & 1 & -1 & 1 \\ 
4 & 0 & -2 & 4
\end{tabular}
\caption{Values of $m$ for the representation $
n_{2}+n_{1}=N=4$, $\eta ^{\prime }=4$, $\eta =\textrm{odd}=3$.}\label{tab_3}
\end{center}
\end{table}
The degeneracies stem from the fact that for a given $N$, there are two values of 
$n_{1}$ and $n_{2}$ satisfying $N = n_{1}+n_{2}$, except for $N = \textrm{even}, n_{1} = n_{2} = N/2$, where the two values merge into a single value. The
dynamic symmetry stems from the simple identity 
\begin{equation}
m^{2}=\left( \frac{n_{2}-n_{1}}{2}\right) ^{2}=\frac{N^{2}}{4}
+(n_{1}^{2}-Nn_{1}),
\end{equation}
which, for $\eta^{\prime }=N$, gives energies counted from the lowest state 
\begin{equation}
    E=n_{1}^{2}-\eta^{\prime }n_{1}. 
\end{equation}

In terms of two boson operators ${\hat b}_{1}$ and ${\hat b}_{2}$, the wave functions of the degenerate states can be written as
\begin{equation}
    |n_1, n_2 \rangle = \frac{1}{\sqrt{n_1! (N-n_1)!} } \left({\hat b}_{2}^{\dag }\right)^{N - n_1} \left({\hat b}_{1}^{\dag }\right)^{n_1} |0 \rangle.
\end{equation}
The notation $|n_1, n_2 \rangle$ can be converted to the usual quasi-spin notation by means of (\ref{eq:jm}) giving 
\begin{equation}
|j, m \rangle = \frac{1}{\sqrt{(j-m)!(j+m)!}}\left({\hat b}_{2}^{\dag }\right)^{j+m}\left({\hat b}_{1}^{\dag }\right)^{j-m} | 0 \rangle.
\end{equation}
Note that the degenerate states are related by the transformation $m\rightarrow -m$ (also related in quantum mechanics to time reversal $T$).
In the two boson construction, it is also possible to associate a parity $P=(-)^{n_{1}}$ to the states. For $N=\textrm{odd}, \eta =\textrm{even}$, the two degenerate
states have opposite parity, while for $N=\textrm{even},\eta =\textrm{odd}$, the degenerate
states have the same parity. Thus, for $N=\textrm{odd}, \eta =\textrm{even}$, the degenerate states
change sign under $PT$ transformation, while for $N=\textrm{even},\eta =\textrm{odd}$ they do
not. All properties of the degenerate points for integer values of $\eta $
can be verified in figure~\ref{fig_1}.

\subsection{Symmetries of the Hamiltonian $\hat{n}(\hat{n}-1)-\protect\xi \hat{P}_{2}$}
\label{subsect_symm_2}

The Hamiltonian
\begin{equation}
\hat{H}_{2}=\frac{\hat{H}}{K}=\hat{n}(\hat{n}-1)-\xi \hat{P}_{2}
\label{H_nodelta}
\end{equation}
is of importance in the theory of quantum phase transitions (QPTs) and of
their associated ESQPTs~\cite{Caprio2008, Cejnar2008, Cejnar2021}. Its structure in terms of elements of the $sp(2,\mathbb{R})\sim su(1,1)$
algebra is
\begin{equation}
\frac{\hat{H}}{K}=\hat{F}_{z}(\hat{F}_{z}-1)-2\xi \hat{F}_{x}~,
\end{equation}
and is therefore in the same universality class of the one-dimensional
vibron model \cite{bookmol, onophd}, $\hat{H}_{\textrm{vibron}}=
\varepsilon \hat{F}_{z}+\delta \hat{F}_{z}^{2}-A\hat{F}_{x}^{2}$ and of the
Lipkin-Meshkov-Glick model~\cite{Lipkin1965}, $\hat{H}_{\textrm{LMG}}=\omega \hat{F}_{z}+v\left(
F_{x}^{2}-F_{y}^{2}\right) $. Its spectrum generating algebra is $sp(2,\mathbb{R})$
with two subalgebras
\begin{equation}
\begin{array}{ccc}
&  & u(1) \\ 
& \nearrow &  \\ 
sp(2,\mathbb{R}) &  &  \\ 
& \searrow &  \\ 
&  & so(1,1)
\end{array}
,
\end{equation}
where $u(1)\doteqdot \hat{a}^{\dag }\hat{a}=\hat{F}_{z}$ and $
so(1,1)\doteqdot \frac{1}{2}\left( \hat{a}^{\dag }\hat{a}^{\dag }+\hat{a}
\hat{a}\right) =\hat{F}_{x}$. Since $sp(2,\mathbb{R})$ is non-compact, in order to
study its symmetries, it is convenient to consider the alternative algebraic
structure of the Heisenberg algebra $u(2)\rightarrow _{c}h(2)$. 

\begin{figure}[t]
    \centering
    \includegraphics[scale=0.8]{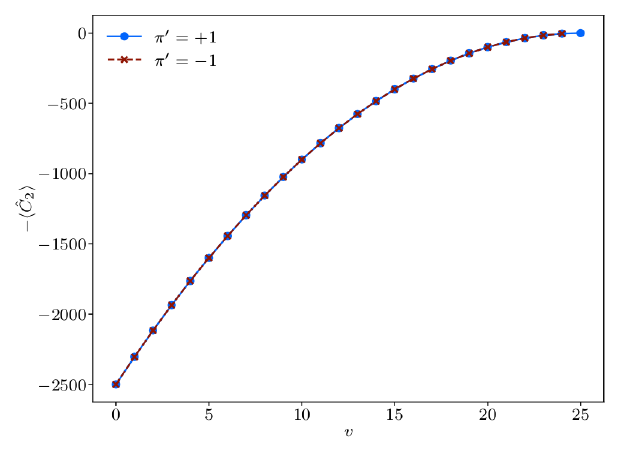}
    \caption{Eigenvalues of the operator $-\hat{C}_{2}$ as a function of $v$ for $
N=50$. Each point is doubly degenerate, $\pi^{\prime }=\pm $, except $v=
N/2=25$.}
    \label{fig_2}
\end{figure}

The algebra
of $u(2)\supset su(2)$ has two subalgebras%
\begin{equation}
\begin{array}{ccc}
&  & u(1) \\ 
& \nearrow &  \\ 
u(2)\supset su(2) &  &  \\ 
& \searrow &  \\ 
&  & so(2)
\end{array}
.
\end{equation}
States are characterized by the quantum numbers
\begin{eqnarray}
u(1) &:&\left\vert \left[ N\right] ,n\right\rangle~,  \nonumber \\
so(2) &:&\left\vert \left[ N\right] ,\sigma \right\rangle~,
\end{eqnarray}
where $n$ are the eigenvalues of $\hat{a}^{\dag }\hat{a}$ $\left(
n=0,1,...,N\right)$ and $\sigma$ those of $\hat{a}^{\dag }\hat{s}+
\hat{s}^{\dag }\hat{a}$ $(\sigma =\pm N,\pm (N-2),\pm (N-4),...,\pm
1 $ or $0$, for $N=\textrm{odd}$ or even). The $\pm $ sign comes from the fact that $so(2) $ is an orthogonal algebra in even dimension \cite{bookalg}. The
notation $\left\vert \left[ N\right] ,\sigma \right\rangle $ can be
converted to the usual $| j,m \rangle $ notation of the quasi-spin algebra $su(2)$ by
\begin{equation}
\left\vert j=\frac{N}{2},m=\frac{\sigma }{2}\right\rangle~,~~m=\pm 
\frac{N}{2},\pm \left(\frac{N}{2}-1\right),\ldots,\pm \frac{1}{2}\textrm {or }0~,
\end{equation}
for $N$ odd or even. The dimension of the representation is 
\begin{equation}
\dim [N]=2j+1=N+1~.
\end{equation}
The value of $j$ is half-integer (integer) for $N$ odd (even). Another notation, used in molecular physics, is \cite{bookmol, onophd} 
\begin{equation}
\left\vert \left[ N\right] ,v^{\pi ^{\prime }}=\frac{N-\sigma }{2}
\right\rangle ~,~~~~\pi ^{\prime }=\pm~,~~v=0,1,\ldots,\frac{N-1}{2}\textrm{ or }\frac{N}{2}~,
\label{eq:v}
\end{equation}
for $N$ odd or even. The quantum number $v$ is called the vibrational quantum number and $\pi^{\prime}$ is the sign of $\sigma$ (or of $m$). In the large system size limit, $N\rightarrow \infty$, the
contracted $so(2)$ operator becomes
\begin{equation}
(\hat{a}^{\dag }\hat{s}+\hat{s}^{\dag }\hat{a})\rightarrow _{c}\frac{(\hat{F}
_{+}+\hat{F}_{-})}{\sqrt{N}}=(\hat{a}^{\dag }+\hat{a})~.
\end{equation}
Another important operator is the quadratic Casimir operator of $so(2)$
\begin{equation}
\hat{C}_{2}=\left( \hat{a}^{\dag }\hat{s}+\hat{s}^{\dag }\hat{a}\right) ^{2}=
\hat{a}^{\dag }\hat{a}^{\dag }\hat{s}\hat{s}+\hat{s}^{\dag }\hat{s}^{\dag }
\hat{a}\hat{a}+2\hat{a}^{\dag }\hat{a}\hat{s}^{\dag }\hat{s}+\hat{a}^{\dag }
\hat{a}+\hat{s}^{\dag }\hat{s}~,
\end{equation}
with eigenvalues 
\begin{equation}
\left\langle \hat{C}_{2}\right\rangle =\sigma ^{2}=4m^{2}~.
\end{equation}
With the vibrational quantum number $v$, the eigenvalues of $\hat{C}
_{2}$ can be written as
\begin{equation}
\left\langle \hat{C}_{2}\right\rangle =N^{2}-4Nv\left( 1-\frac{v}{N}
\right) , ~~~v=0,1,\ldots,\frac{N-1}{2}\textrm{ or }\frac{N}{2}~, 
\end{equation}
for $N$ odd or even and $\pi^{\prime }=\pm$. 
The eigenvalues of $\hat{C}_{2}$ are doubly degenerate $\pi ^{\prime }=\pm $, except for $\sigma =0$ ($v=N/2$, even $N$) which is singly degenerate.
Introducing the pairing operator of $su(2)$
\begin{equation}
\hat{P}_{2}^{\prime }=\hat{a}^{\dag }\hat{a}^{\dag }\hat{a}\hat{a}+\hat{s}
^{\dag }\hat{s}^{\dag }\hat{s}\hat{s}-\hat{a}^{\dag }\hat{a}^{\dag }\hat{s}
\hat{s}-\hat{s}^{\dag }\hat{s}^{\dag }\hat{a}\hat{a}=N^{2}-\hat{C}_{2}~,
\end{equation}
one has
\begin{equation}
\left\langle \hat{P}_{2}^{\prime }\right\rangle = N^{2}-\sigma ^{2}=4Nv\left(1-
\frac{v}{N}\right), ~~~v=0,1,\ldots,\frac{N-1}{2}\textrm{ or }\frac{N}{2}~,
\label{eq:P2prime}
\end{equation}
for $N$ odd or even and $\pi^{\prime }=\pm$. The eigenvalues of $-\hat{C}_{2}$ are shown in figure~\ref{fig_2}.

The contracted form of the operator $\hat{C}_{2}$ is 
\begin{equation}
\hat{C}_{2}\rightarrow _{c}N(\hat{a}^{\dag }\hat{a}^{\dag }+\hat{a}\hat{a})+2%
\hat{n}(N-\hat{n})+N~,
\end{equation}
where $\hat{n}_{s}=\hat{s}^{\dag }\hat{s}$ has been replaced by $N-\hat{n}$.
Unfortunately, because of the additional terms, the eigenvalues of the
pairing operator of $sp(2,\mathbb{R})$, $\hat{P}_{2}=\hat{a}^{\dag }\hat{a}^{\dag }+%
\hat{a}\hat{a}$, cannot be obtained simply from those of the quadratic
Casimir operator of $so(2)$ and must be calculated numerically.

\begin{figure}[h]
    \centering
    \includegraphics[scale=0.8]{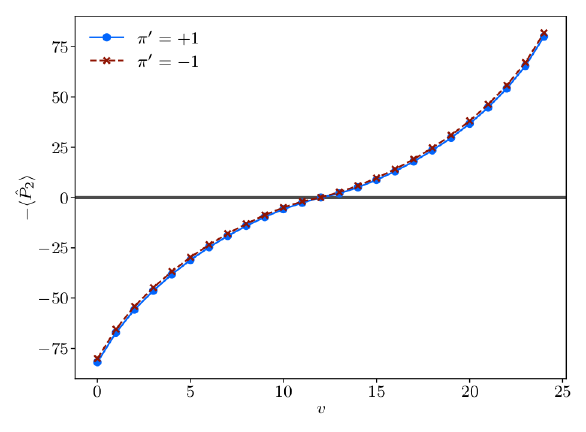}
    \caption{Eigenvalues of the operator $-\hat{P}_{2}$ as a function of $v$ for $N=50$.}
    \label{fig_3}
\end{figure}

Going from $su(2)$ to $sp(2,\mathbb{R})$, while the $u(1)$ classification $ |  \left[ N\right] ,n  \rangle $ remains the same, the classification $ | \left[ N\right] ,\sigma \rangle$ in terms of a quasi-spin $j$ and component $m$ needs to be modified, since there is a doubling of representations \cite{Wybourne1974}. To this end, we consider the eigenvalues of
the operator 
\begin{equation}
-\hat{P}_{2}=-\left( \hat{a}^{\dag 2}+\hat{a}^{2}\right) =-2\hat{F}_{x}~.
\end{equation}
Introducing the parity $\pi =\left( -\right) ^{n}$, for a given $N$, there
are two representations, one with even parity $\pi =+$ and one with odd
parity $\pi =-$, with values of $j$ given by
\begin{equation}
\begin{array}{ccccc}
N=4\nu &;& j=\frac{N}{4},~\pi =+ &; & j=\frac{N}{4}-\frac{1}{2},~\pi =- \\ 
N=4\nu +2 &;& j=\frac{N}{4},~\pi =+ &;& j=\frac{N}{4}-\frac{1}{2},~\pi =- \\ 
N=4\nu +1 &;& j=\frac{N-1}{4},~\pi =+ &;& j=\frac{N-1}{4},~\pi =- \\ 
N=4\nu +3 &;& j=\frac{N-1}{4},~\pi =+ &;& j=\frac{N-1}{4},~\pi =-
\end{array}
\end{equation}
and $\nu =1,2,3,...$. The values of $m$ are 
\begin{equation}
m=j,j-1,...,-(j-1),-j=j-v~,
\end{equation}
with $v$ given by 
\begin{equation}
\begin{array}{ccccc}
N=\textrm{even} &;& v=0,1,...,\frac{N}{2}~,~~\pi =+ &;& v=0,1,...,(\frac{N}{2}-1)~,~~\pi =- ~,\\ 
N=\textrm{odd} &;& v=0,1,...,\frac{N-1}{2}~,~~\pi =+ &;& v=0,1,...,\frac{N-1}{2}~,~~\pi =- ~.
\end{array}
\end{equation}
For example, for $N=50$, the even parity states are classified by the
representation $j=\frac{25}{2},$ $-\frac{25}{2}\leq m\leq +\frac{25}{2}$,
with a total number of states $2j+1=26$. The odd parity states are classified by the
representation $j=\frac{24}{2}=12,$ $-12\leq m\leq +12$, with a total number
of states $2j+1=25$. For each representation, the eigenvalues come in pairs,
corresponding to positive and negative values of $m=\pm j,\pm (j-1), \ldots,\pm 1/2$ or $0$ for $j$ half-integer or integer ($\pi^{\prime }$, $m\rightarrow -m$), except for $m=0$. The results of a numerical
diagonalization are shown in figure~\ref{fig_3} for $N=50$. For each parity $\pi =\pm $,
there are two branches with $\pi ^{\prime }=\pm $. Note that $\pi \neq \pi^{\prime }$.

In the limit $N\rightarrow \infty $, the spectrum of the operator $-\hat{P}
_{2}$ is a straight line extending from $-N$ to $+N$. The line is doubled,
one for each parity $\pi =\pm $. This property stems from the fact that the
operator $\hat{F}_{x}$ changes $n$ by $\pm 2$ units and thus conserves parity. The
spectrum of energies can be written as 
\begin{equation}
E=-2m~.
\end{equation}
The spectrum extends to $\pm \infty $ due to the non-compact nature of $sp(2,\mathbb{R})$, the representations of which are discrete but infinite
dimensional, $-\infty <m<+\infty $. Note that in the $N\rightarrow \infty $
limit, the two representations, $j_{even}$ and $j_{odd}$, which form the two
components of the $sp(2,\mathbb{R})$ representation, become degenerate. However, as
seen from figure~\ref{fig_3}, the convergence to the asymptotic limit is very slow, and it is far from reached at $N=50$.

\begin{figure}[h]
    \centering
    \includegraphics[scale=0.8]{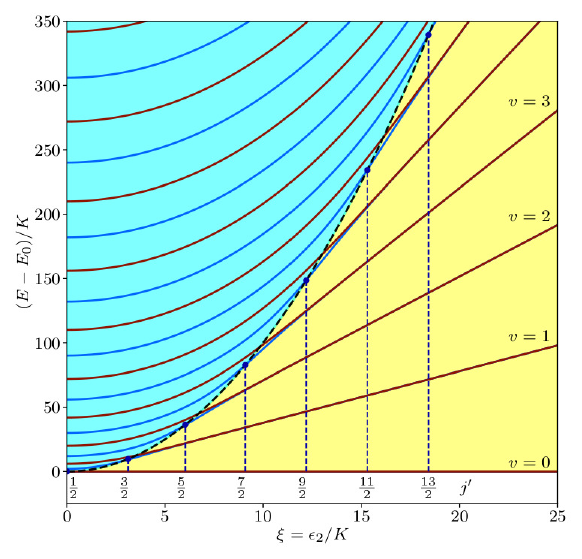}
    \caption{Excitation energy of Hamiltonian (\ref{H_nodelta}) as a function of the control parameter $\xi$. Quasi-spin $| j^{\prime },m^{\prime } \rangle $
classification of the degenerate states at the point of maximal rate of
approach, $\xi_{c}=\pi v$ is shown. The separatrix, marked with a dashed black line, is $E_{s}=\xi^{2}$. The size of the Fock space is truncated at $N = 800$. States to the left of the separatrix are singly-degenerate with positive (blue) and negative (red) parity. States to the right of the separatrix are labeled by the number $v=0,1,2, \ldots$ of equation (\ref{eq:v}) and are doubly degenerate with parity $\pm.$}
    \label{fig_4}
\end{figure}

Consider now the Hamiltonian (\ref{H_nodelta}). The spectrum of this Hamiltonian,
calculated numerically in the $u(1)$ basis with $n_{\max }\equiv N=800$, is
shown in figure~\ref{fig_4}. The
value of $N=800$ is chosen here in order that the eigenvalues in the range $0\leq \xi \leq 25$ of the figure are well converged. While the convergence of the eigenvalues of the pairing term $\hat{P}_{2}$ is very slow due to its non-compacteness, the convergence of the Hamiltonian (\ref{H_nodelta}) is faster due to the presence of the Kerr term $\hat{n} (\hat{n}-1)$ which increases as $n^2$ for large $n$. The spectrum exhibits an ESQPT~\cite{Chavez2022} similar to that encountered in the one-dimensional vibron model and the Lipkin-Meshkov-Glick model~\cite{Cejnar2021,Heiss2005,Santos2016}. It is
divided into two parts (phases) with separatrix $E_{s}$. To the left of the
separatrix, states are singly degenerate with $u(1)$ symmetry. To the right,
states are doubly degenerate with $so(2)$ symmetry. The degenerate states
have opposite parity $\pi$. The classification of states in terms of a
quasi-spin $\left\vert j^{\prime },m^{\prime }\right\rangle $ is, however, as
discussed in the paragraphs above, rather complicated. At values of $\xi
=4v=0,4,8,12,\ldots$ states can be classified by a quasi-spin $\left\vert
j^{\prime },m^{\prime }\right\rangle $ with $j^{\prime }=\xi/4+1/2=1/2,3/2,5/2,...$ and $-j^{\prime }\leq m^{\prime }\leq +j^{\prime }$,
where we have used $j^{\prime },m^{\prime }$ instead of $j,m$ to
emphasize that the quasi-spin here is not the same of the previous
subsection. This classification is also valid at values of $\xi_c =(\pi /4)4v$,
which are the values of the critical points as obtained from the maximal
rate of approach~\cite{Frattini2022} and shown with circles in figure~\ref{fig_4}.

\begin{figure}[h]
    \centering
    \includegraphics[scale=0.8]{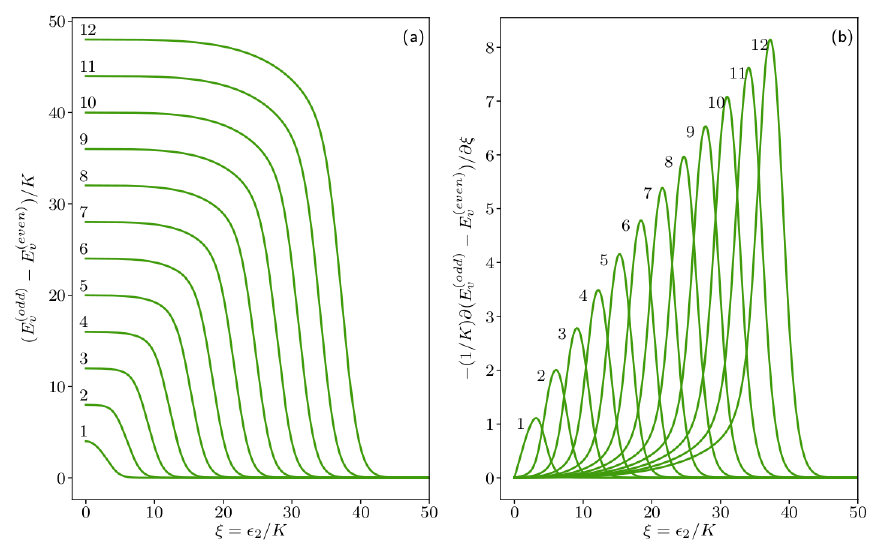}
    \caption{Panel (a): The gap $(E_{v}^{(odd)}-E_{v}^{(even)})/K$ as a function of $\xi $ for $v=1,\ldots,12$. Panel (b): The derivative of the gap $-\frac{1}{K} \partial(E_{v}^{(odd)}-E_{v}^{(even)})/\partial \xi $ as a function of $\xi $ for $v=1,\ldots,12$. The critical value $\xi_{c}$ is the value of $\xi $ at the
maximum of the derivative~\cite{Frattini2022}. The gaps go to zero as the separatrix is crossed for each $v$ as a function of $\xi$. To the right of the separatrix, states are doubly degenerate with $E_{v}^{(odd)}=E_{v}^{(even)}$ (parity $\pm$).}
    \label{fig_5}
\end{figure}

\begin{figure}[h]
    \centering
    \includegraphics[scale=0.8]{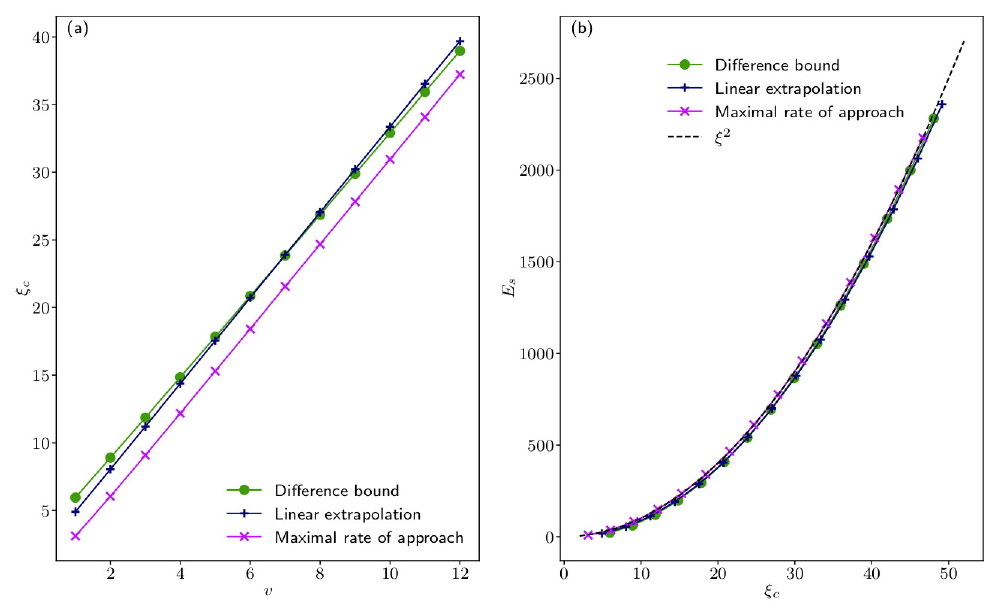}
    \caption{Panel (a): The values of the critical point $\xi _{c}$ as a function of $v$ as obtained by the three methods: maximal rate of approach (cross) [1],
linear extrapolation (plus), and difference bound (circle). Panel (b): The
values of the separatrix $E_{s}$ as a function of $\xi $ as obtained by the
three methods of panel (a). The semiclassical result \cite{Chavez2022} is shown with a dashed
line.}
    \label{fig_6}
\end{figure}

In the large $N$ limit, the separatrix is $E_{s}=\xi^{2}$, the energy of
the states to the right of the separatrix is $E_{v}=4\xi v$ ($v=0,1,2,3...$)
and the critical value obtained by the condition $E_{s}=E_{v}$ is $\xi_{c}=4v$. The values of $E_{v}$ to which the numerical calculation converges cannot be obtained in explicit analytic form. By analogy with (\ref{eq:P2prime}), which applies to the compact version of the operator, $\hat{P}'_2$, we suggest an approximate expression for the energy of the states to the right of the separatrix to be
\begin{equation}
E_{v}=4\xi v\left( 1-\frac{v}{N_{\textrm{eff}}}\right) ~,~~v=0, 1, 2, 3, \ldots
\end{equation}
where $N_{\textrm{eff}}=N/2$ and $N$ is the value $n_{\max }$ of states kept in the numerical calculation. The determination of the critical value $\xi_{c}$
depends on its definition. 
In Ref.~\cite{Frattini2022},
the critical value is defined as the point of maximal rate of approach determined 
by the inflection point in the energy gaps, as illustrated in figure~\ref{fig_5}. The value so determined is $\xi_{c}=0.775(4v)\cong \pi v$~\cite{Frattini2022}. However, at this point, the gap is still
large. Another possible determination is by a linear extrapolation of the gaps $(E_{odd}-E_{even})$.
This determination is closer to the expected values of the ESQPT. Finally,
another determination is the location at which the energy difference $(E_{odd}-E_{even})$ is less than a given fraction of the energy semi-sum $
(E_{odd}-E_{even})/2$. We use here $\left(E_{odd}-E_{even}\right) \leq
0.005(E_{odd}+E_{even})/2$. The values determined by these three methods are
shown in figure~\ref{fig_6}(a). 

All three results produce similar results. Particularly interesting is the
result of the maximal rate of approach, which is a straight line with slope $\pi$, as expected in a semi-classical approximation to the Hamiltonian $\hat{H}$~\cite{Frattini2022}. However, this result is not accurate for small $v$, since at this
point the gap is still large. For small $v$, the critical value that best
describes the merging of the two energies $E_{odd}-E_{even}$ (the so-called
kissing point~\cite{Frattini2022}) is the linear extrapolation. The values of $E_{c}$ at the
critical point $\xi _{c}$ determine then the separatrix $E_{s}$. The values
so determined are shown in figure~\ref{fig_6}(b). Again here all three methods produce
similar results all of which are very close to the semi-classical expression 
$E_{s}=\xi ^{2}$ \cite{Chavez2022}.

\subsection{Symmetries of the Hamiltonian $-\protect\eta \hat{n}+\hat{n}(\hat{n}-1)-\protect\xi \hat{P}_{2}$}
\label{subsect_symm_3}

\begin{figure}[h]
    \centering
    \includegraphics[scale=0.8]{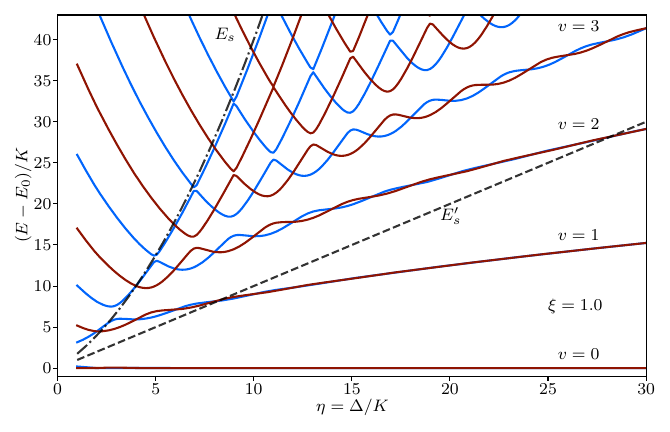}
    \caption{Spectrum of the Hamiltonian (\ref{Kerr_Squeezed}) as a function of $\eta $ for $\xi =1$. The approximate
separatrices $E_{s}=\eta/2+\eta^{2}/4+\eta \xi $ and $E_{s}^{\prime }=\eta \xi $ are shown as black dash-dotted and dashed lines, respectively. To the left of the separatrix $E_{s}$, Phase I, states are singly degenerate with positive (red) and negative (blue) parity. In between the separatrices $E_{s}$ and $E'_{s}$, Phase II, braiding occurs. To the right of the separatrix, Phase III, states are doubly degenerate with parity $\pm$. They are labeled by the quantum number $v=0,1,2,\dots$ of equation~(\ref{eq:v}). The calculations have been carried out with a Fock basis with $N = 800$. 
}
    \label{fig_7ab}
\end{figure}

The spectrum of the Hamiltonian (\ref{Kerr_Squeezed}), $\hat{H}/K=-\eta \hat{n}+\hat{n}(\hat{n}-1)-\xi \hat{P}_{2}$,
 as a function of $\eta $ for a fixed value of $\xi =1$ is shown in figure~\ref{fig_7ab}(a).
The spectrum is now separated into three phases.

Phase 1 is separated from phase 2 by a separatrix approximately given by 
\begin{equation}
E_{s}=\frac{\eta }{2}+\frac{\eta ^{2}}{4}+\eta \xi~.
\label{separatrix_1}
\end{equation}
To the left of the separatrix, given by (\ref{separatrix_1}), states are still singly
degenerate with $u(1)$ symmetry. To the right of this separatrix, states
split into phases 2 and 3, with another separatrix, the energy of which, $E_{s}^{\prime }$, is approximately given by 
\begin{equation}
E_{s}^{\prime }=\eta \xi~.
\label{separatrix_2}
\end{equation}

In the intermediate phase 2, to the left of the new separatrix (\ref{separatrix_2}), states resemble those of
phase 2 of the Hamiltonian $\hat{H}_{1}=-\eta \hat{n}+\hat{n}(\hat{n}-1)$ in figure~\ref{fig_1},
with braiding which decreases as $\eta $ increases. However, while at even
values of $\eta =2,4,6,\ldots$ states maintain the double degeneracy of the $so(2)$ symmetry of figure~\ref{fig_1}, at odd values of $\eta =1,3,5,\ldots$ the degeneracy
is lifted and replaced by a new $so(2)$ symmetry, as will be discussed in the next section. To be precise, each
 half-integer $j$ quasi-spin representation that occurs at even
values, $\eta =2,4,6, \ldots$, is simply moved up. The crossings are allowed since the
degenerate states, composed by one positive and one negative parity
state ($\pi =\pm$) have different symmetries. Each integer $j$ quasi-spin representation, which
occurs for odd values $\eta =1,3,5,\ldots$, is instead modified. In this case states have the same parity, the crossing is forbidden and the degeneracy is lifted. The different character of the crossings at even and odd values of the $\eta$ control parameter have a strong impact on the system dynamics and, in particular, on the possibility of tunneling between different regions of the system's phase space \cite{Prado2023}. Similar effects have been recently uncovered in the Lipkin-Meshkov-Glick model ESQPT \cite{Nader2021}.

In phase 3, which lies to the right
of the new separatrix (\ref{separatrix_2}), states resemble those of phase 2 of
Hamiltonian $\hat{H}_{2}=\hat{n}(\hat{n}-1)-\xi \hat{P}_{2}$ in figure~\ref{fig_4}. States can be
classified by a vibrational quantum number $v=0,1,2,\ldots$. (or by a
quasi-spin $\left\vert j^{\prime },m^{\prime }\right\rangle $ with 
$j^{\prime }=$ half-integer, $m^{\prime }=\pm 1/2,\pm 3/2,...$). 

As it will be shown in the next section, the location of the degeneracies for even $\eta
=2,4,6, \ldots$ in the intermediate phase 2 remains the same even for
non-perturbative values of $\xi$, due to the fact that the quasi-spin
quantum numbers retain the same half-integer values, while for odd 
$\eta =1,3,5,\ldots$, the degeneracies change from those of integer $j$
to those of half-integer $j$.  

\section{Other Hamiltonians and the stability of the solutions of the squeeze-driven Kerr Hamiltonian}
\label{sect_stability}

In view of the fact that additional Hamiltonians can be, in principle, engineered and experimentally studied \cite{Frattini2022, xiao2023diagrammatic}, we consider in this section the effect of adding parametric terms to the Kerr nonlinearity. We begin by investigating 
\begin{equation}
\frac{\hat{H}}{K}=-\eta \hat{n}+\hat{n}(\hat{n}-1)-\xi_{i}\hat{P}_{i}~,
\end{equation}
where $\hat{P}_{i}$ is a generic term and $\xi_{i}$ its strength.

\begin{figure}[h]
    \centering
    \includegraphics[scale=0.8]{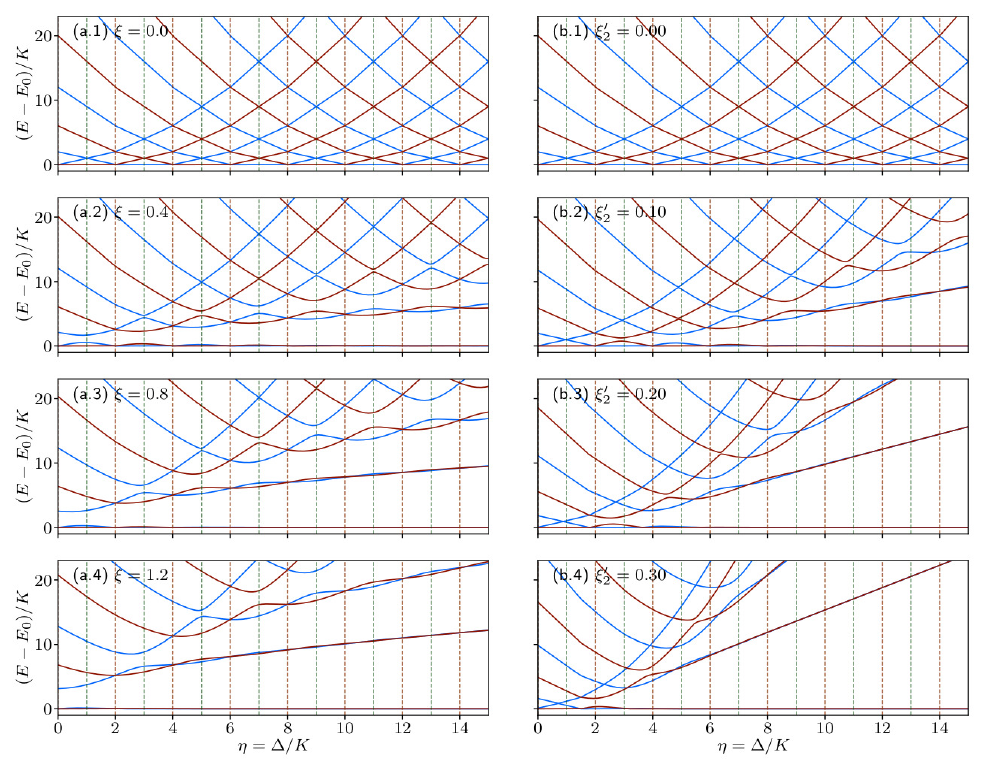}
    \caption{Panels (a.1-4): Spectrum of the Hamiltonian (\ref{Ham_P2}) as a function of $\eta $ for different values of $\xi$. Panels (b.1-4): Spectrum of the Hamiltonian (\ref{Ham_nP}) as a function of $\eta $ for
different values of $\xi_{2}^{\prime }$. All panels: The
integer values of $\eta $ are shown by vertical dotted lines. The blue (red) color denotes
even (odd) parity states.}
    \label{fig_8}
\end{figure}

Consider first the addition of the term $-\xi \hat{P}_{2}$, which takes the
Hamiltonian of the Kerr oscillator (\ref{Kerr_no_squeezing}) to that of the squeeze-driven Kerr
oscillator (\ref{Kerr_Squeezed}), copied here once again,
\begin{equation}
\frac{\hat{H}}{K}=-\eta \hat{n}+\hat{n}(\hat{n}-1)-\xi \hat{P}_{2}~.
\label{Ham_P2}
\end{equation}
Hamiltonian (\ref{Kerr_no_squeezing}) has, as discussed in subsection~\ref{subsect_symm_1}, a quasi-spin symmetry with states characterized by the quasi-spin quantum
numbers $| j,m \rangle $, with integer $j$ at $\eta=1,3,5 \ldots$ and half-integer $j$ at $\eta =2,4,6, \ldots$. A remarkable property of the squeeze-driven Kerr Hamiltonian (\ref{Ham_P2}) is that the quasi-spin representations 
$| j,m \rangle $ with $j=$ half-integer (even $\eta =2,4,6,\ldots$) are not altered by the term $-\xi \hat{P}_{2}$, that is, the
degeneracies remain at the same points $\eta =2,4,6...$ but at larger energy values. This is clearly seen in figures~\ref{fig_8}(a.1)-(a.4), where the 
spectrum of the Hamiltonian (\ref{Ham_P2}) is shown as a function of $\eta $ for different values of $\xi $. This property, first found in~\cite{Frattini2022,Venkatraman2022_Exp,VenkatramanThesis}, is of great importance for possible applications of KPO to quantum computing, and is related to the newly discovered quasi-spin symmetry which occurs for integer values of $\eta=\Delta/K$.

\begin{figure}[h]
    \centering
    \includegraphics[scale=0.8]{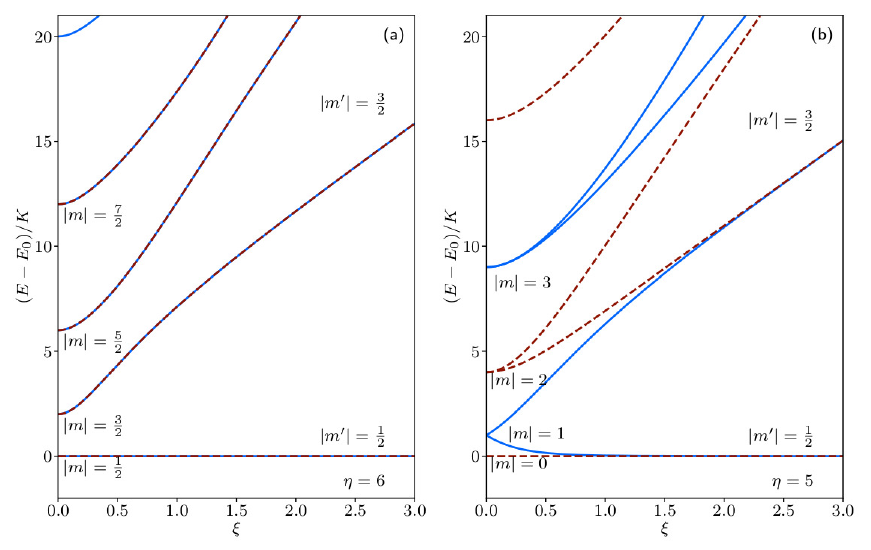}
    \caption{Panel (a): Behavior of the different components of the quasi-spin $j=7/2, \eta =6$ as a function of $\xi $. The degeneracies persist even for
non-perturbative values of $\xi$. Panel (b): Same as (a) but for $j=3,\eta =5$.
The degeneracies in panel (b) are split. In the asymptotic limit, not yet reached in
the figure, the degeneracies become those of $j^{\prime }=7/2$. See also \cite{Zhang2017}.
Both panels: solid blue lines are for even parity states and dashed red lines for odd parity states.
}
    \label{fig_9}
\end{figure}

The behavior of the
different $m$ components of the quasi-spin $j$ as a function of $\xi$ is
shown in figure~\ref{fig_9}(a) for the representation $j=7/2$ ($\eta =6$). At $\xi =0$,
the energies are given by (\ref{spectrum_nodelta_1}), $E=m^{2}-1/4$.
On the contrary, the quasi-spin representations with integer $j$ (odd $\eta =1, 3, 5, \ldots$) are affected in such a way that the degeneracies
change from those of integer $j$, with $m=0,\pm 1,\pm 2, \ldots$, to those of half-integer $j$, with $m=\pm 1/2,\pm 3/2, \ldots$. The behavior of the
different $m$ components of the quasi-spin $j$ as a function of $\xi$ in the odd $\eta$ case is
shown in figure~\ref{fig_9}(b) for the representation $j=3$ ($\eta =5$). At $\xi =0$, the 
energies are $E=m^{2}$, as given in (\ref{spectrum_nodelta_2}). These properties persist even for
non-perturbative values of $\xi$.

Going to third and fourth orders in the boson expansion of the Hamiltonian [see equations (\ref{order_3}) and (\ref{order_4})], we observe that some of the additional terms are a
renormalization of lower order terms, $\Delta ^{(3)},K^{(3)},\varepsilon
_{2}^{(3)},\Delta ^{(4)},K^{(4)},\lambda ^{(4)}$, but two new terms appear,
one of order 3, 
\begin{equation}
\varepsilon _{2}^{\prime }(\hat{n}\cdot \hat{P}_{2})\equiv \varepsilon
_{2}^{\prime }\left( \hat{a}^{\dag 2}(\hat{a}^{\dag }\hat{a})+(\hat{a}^{\dag
}\hat{a})\hat{a}^{2}\right)~,
\end{equation}
where the dot indicates normal ordering, and one of order 4,
\begin{equation}
\varepsilon _{4}^{(4)}\hat{P}_{4}\equiv \varepsilon _{4}^{(4)}(\hat{a}^{\dag
4}+\hat{a}^{4})~.
\end{equation}

\begin{figure}[h]
    \centering
    \includegraphics[scale=0.8]{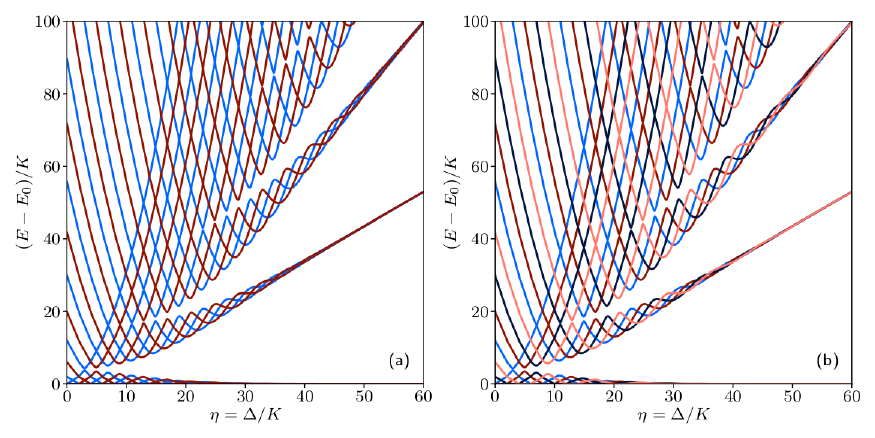}
    \caption{ Panel (a): Spectrum of the Hamiltonian (\ref{Ham_P4}) for $\xi _{4}=0.05$. The blue color denotes
positive parity states and the red color negative parity states with $\textrm{mod}\,2\oplus \textrm{mod}\,2$ coloring. Panel (b):
Spectrum of the same Hamiltonian as a function of $\eta $ with $\textrm{mod}\,4$ coloring, with red tones for odd parity and blue tones for even parity states. In all panels, calculations have been carried out with a Fock basis truncated at $N=800$. }
    \label{fig_10ab}
\end{figure}
\begin{figure}[h]
    \centering
    \includegraphics[scale=0.8]{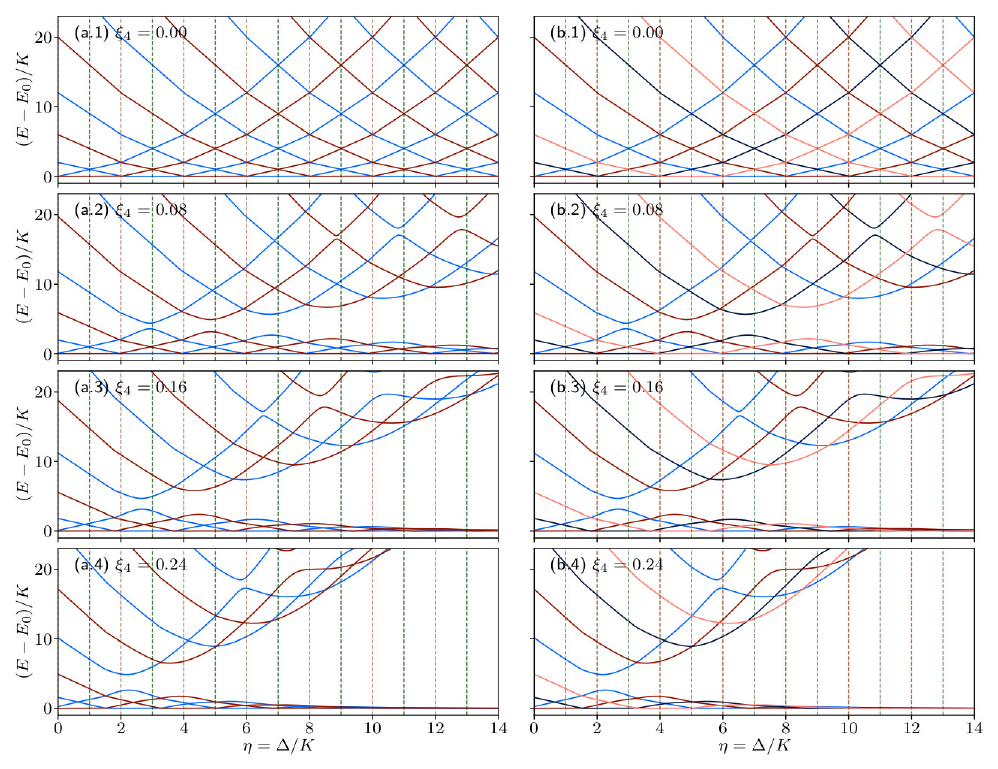}
    \caption{ Spectrum of the Hamiltonian  (\ref{Ham_P4}) as a function of $\eta $ for
different values of $\xi_{4}$. Panels (a.1-4): States identified with a $\textrm{mod}\,2\oplus \textrm{mod}\,2$ coloring. The blue color denotes
positive parity states and the red color negative parity states. Panels (b.1-4):
States identified  with a $\textrm{mod}\,4$ coloring scheme. In all panels, calculations have been carried out with a Fock basis truncated at $N=800$. }
    \label{fig_11}
\end{figure}

The spectrum of the Hamiltonian 
\begin{equation}
\frac{\hat{H}}{K}=-\eta \hat{n}+\hat{n}(\hat{n}-1)-\xi _{2}^{\prime }\hat{n} \cdot \hat{P}
\label{Ham_nP}
\end{equation}
as a function of $\eta $, for different values of $\xi_{2}^{\prime}$,
is shown in figures~\ref{fig_8}(b.1)-(b.4).  It is similar to that of the $\hat{P}_{2}$ term with a
two-fold degeneracy in the asymptotic limit of $\xi_{2}^{\prime}$ large and
requires no further comment.

The spectrum of the Hamiltonian 
\begin{equation}
\frac{\hat{H}}{K}=-\eta \hat{n}+\hat{n}(\hat{n}-1)-\xi _{4}\hat{P}_{4}
\label{Ham_P4}
\end{equation}
is shown in the panels of figure~\ref{fig_10ab} for $\xi _{4}=0.05$. This Hamiltonian is the same that was studied in Ref.~\cite{Kwon2022}, where  a promising quantum error correction scheme was proposed. The spectrum of this Hamiltonian has remarkable new
features when compared with that of the $\hat{P}_{2}$ term in (\ref{Ham_P2}) depicted in figure~\ref{fig_7ab}(a);
the most notable being that, in the asymptotic limit, the eigenstates of the Hamiltonian (\ref{Ham_P4}) have four-fold degeneracy. To clarify this situation, we show in 
figure~\ref{fig_10ab}(a) the eigenvalues as a function of $\eta $ for $\xi_{4}=0.05$, colored by parity. We see here crossings of states both with
opposite parity, $\pi =+,-$, and with the same parity, $\pi =+,$ $+$ and $\pi =-,- $, and also avoided crossings of states with the same parity. This
property is due to the fact that the $\hat{P}_{4}$ term couples states with
oscillator quantum number $n$ differing by four units, i.e.~$n\, \textrm{mod}\, 4$.
This property is evinced in figure~\ref{fig_10ab}(b), where now the states are colored with $\textrm{mod}\, 4$. In other words, while in  figure~\ref{fig_10ab}(a) the coloring is that of two copies of
the cyclic group $C_{2}\equiv \Pi $, in figure~\ref{fig_10ab}(b) the coloring is that of the cyclic group, $C_{4}$, where $C_{\nu }$ is the cyclic group of order $\nu $.
This is a special property of the group $C_{4}$, which can be split into $C_{2}\oplus C_{2}$. The spectrum of the Hamiltonian (\ref{Ham_P4}) as a function of $\eta$ for different values of $\xi_4$ is shown in figure~\ref{fig_11}, with states in panels (a.1-4) depicted using the $\textrm{mod}\,2\oplus \textrm{mod}\,2$ coloring scheme and in panels (b.1-4) using the  $\textrm{mod}\, 4$ scheme. Note that this $\hat P_4$ Hamiltonian term can be implemented in  experiments like those in \cite{Frattini2022,Venkatraman2022_Exp} by further engineering the hardware \cite{miano2023hamiltonian} and driving conditions \cite{xiao2023diagrammatic,Kwon2022}.

\begin{figure}[h]
    \centering
    \includegraphics[scale=0.8]{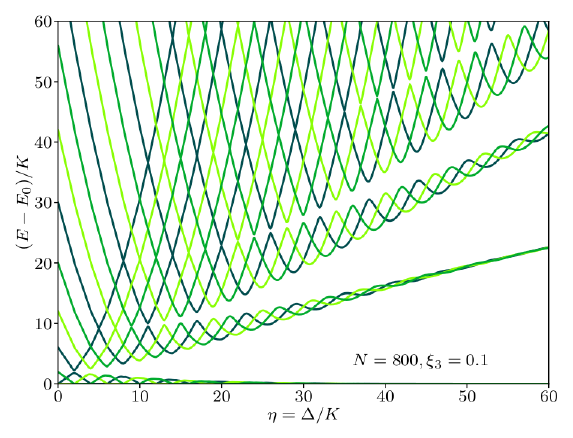}
    \caption{Spectrum of the Hamiltonian (\ref{Eq:P3}) for $\xi _{3}=0.1$. The different shades of green
denote states within a $\textrm{mod}\,3$ coloring scheme. The Fock space is truncated at $N=800$.}
    \label{fig_12}
\end{figure}

\begin{figure}[h]
    \centering
    \includegraphics[scale=0.8]{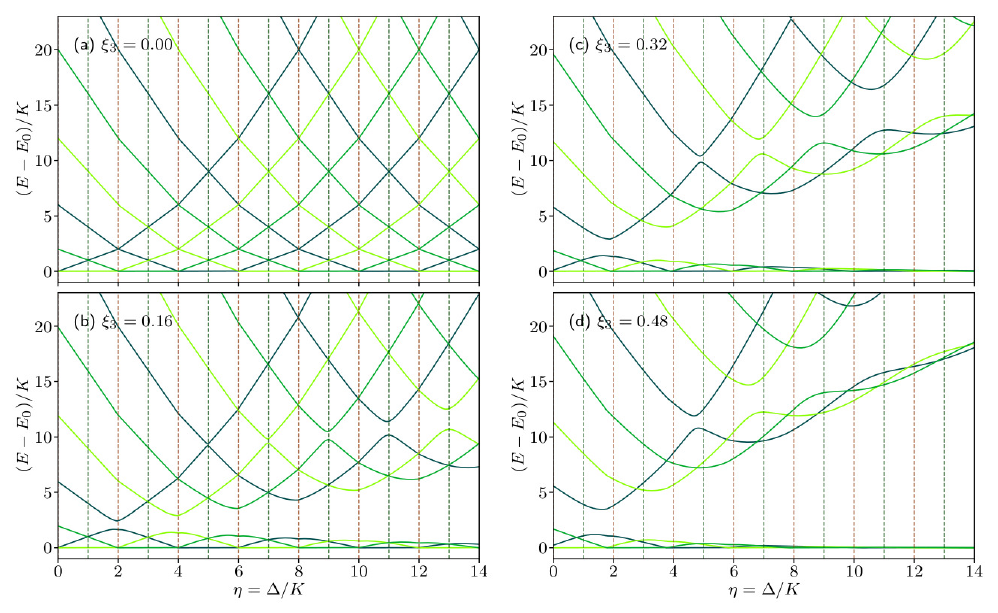}
    \caption{Spectrum of Hamiltonian (\ref{Eq:P3}) as a
function of $\eta $ for different values of $\xi _{3}$ with a $\textrm{mod}\,3$
coloring scheme. The Fock space is truncated at $N=800$.}
    \label{fig_13}
\end{figure}

For purposes of studying the stability of the solutions of the Hamiltonian
(\ref{Kerr_no_squeezing}), it is also of interest to consider the effect of other
perturbations, in addition to those contained in the Hamiltonian of section~\ref{sec_intro}.
Particularly interesting is the term of order 3, which can be experimentally implemented via the correct driving condition \cite{xiao2023diagrammatic,Zhang2017},
\begin{equation}
\varepsilon _{3}\hat{P}_{3}=\varepsilon _{3}\left( \hat{a}^{\dag 3}+\hat{a}^{3}\right)~.
\end{equation}
The spectrum of the Hamiltonian 
\begin{equation}
\frac{\hat{H}}{K}=-\eta \hat{n}+\hat{n}(\hat{n}-1)-\xi_{3}\hat{P}_{3}
\label{Eq:P3}
\end{equation}
is shown in figure~\ref{fig_12} for $\xi_{3}=0.1$ and in figure~\ref{fig_13} as a function of $\eta$ for different values of $\xi_{3}$. This spectrum, first studied in~\cite{VenkatramanThesis}, has also some remarkable
properties, since in the asymptotic limit, the states have a three-fold
degeneracy and are representations of the cyclic group $C_{3}$. Also, parity
here is not a good quantum number, since the $\hat{P}_{3}$ term couples
states with $n$ differing by three units, i.e.~$n\, \textrm{mod}\, 3$. The coloring with different shades of green in figures \ref{fig_12}  and \ref{fig_13} reflects this property.
\begin{figure}[h]
    \centering
    \includegraphics[scale=0.8]{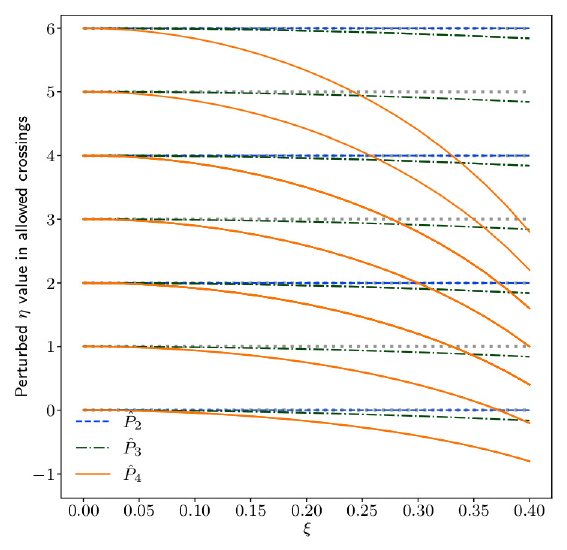}
    \caption{Behavior of the location of the degeneracies $\pm m$  as a function of the strength $\xi $ for the
representations $j=1/2$, $3/2$, $5/2$, and $7/2$ ($\eta =0, 2, 4, 6$) for $\hat{P}_{2}$, $\hat{P}_{3}$, and $\hat{P}_{4} $ and the representations $j=1, 2$, and $3$ ($\eta =1, 3, 5$) for $\hat{P}_{3}$ and $\hat{P}_{4}$.
}
    \label{fig_14}
\end{figure}

 It is remarkable that the quasi-spin symmetry of Hamiltonian (\ref{Kerr_no_squeezing}) is robust for the term $\hat{P}_{2}$: the value of $\eta $
remains at even $\eta=2,4,6,...$ for crossings of states with different
parity and at odd $\eta =1,3,5,...$ for avoided crossings of states with the
same parity. For the other terms, the value of $\eta$ changes  and other
crossings and avoided crossings occur. The change of the value of $\eta $
from even $\eta=0,2,4,6$ as a function of the coupling strength $\xi $ for
the perturbations $\hat{P}_{2}$, $\hat{P}_{3}$, and $\hat{P}_{4}$ is shown in figure~\ref{fig_14}. One can see from this figure that for $\hat{P}_{2}$, no change occurs, even
for large (non-perturbative) values of the perturbation strength $\xi$. The change is small and independent of $\eta$
for $\hat{P}_{3}$ and considerably larger and dependent on $\eta$ for $\hat{P}_{4}$. As shown in figure~\ref{fig_14}, the $\hat{P}_{3}$ and $\hat{P}_{4}$ perturbations have also allowed crossings for odd $\eta$ values, with the same dependence than in the even $\eta$ cases: small and $\eta$ independent variations in the $\hat{P}_{3}$ case and larger and $\eta$-dependent variations in the $\hat{P}_{4}$ case. These results have important implications for Hamiltonian stabilization of bosonic codes for quantum information \cite{Frattini2022,Venkatraman2022_Exp,grimm2020}.

Another interesting result is that in the asymptotic limit of large coupling 
$\xi $, the degeneracies change from the two-fold degeneracy of $\hat{P}_{2}$
to a three-fold degeneracy in $\hat{P}_{3}$ and a four-fold degeneracy in $%
\hat{P}_{4}$. A full study of the algebraic structure of the $\hat{P}_{3}$
and $\hat{P}_{4}$ terms and their associated symmetries remains to be done,
especially in relation to their braiding properties shown in figure~\ref{fig_10ab} and figure~\ref{fig_12}.

\section{Summary and conclusions}
\label{sect_summary}

In this article, we have investigated the symmetries of the squeeze-driven Kerr
oscillator, discovered a hitherto unknown quasi-spin symmetry of the
Hamiltonian $\hat{H}=-\Delta \hat{n}+K\hat{n}(\hat{n}-1)$, and shown that
solutions at even values of the ratio $\eta =\Delta$(detuning)$/K$(Kerr
coefficient) are very stable to perturbations induced by the ratio $\xi
=\varepsilon _{2}$(squeezing amplitude)$/K$(Kerr coefficient), and
moderately stable to perturbations induced by high order terms in the boson
expansion of the Hamiltonian. This result has major implications for the use
of the squeeze-driven Kerr oscillator in quantum computing. In particular, the discovery of the quasi-spin symmetry of the Kerr-Hamiltonian may have major implications when going from a KPO (Kerr Parametric Oscillator) to an OPO (Open Parametric Oscillator), that is from the solutions of the
Hamiltonian operator to the solutions of the Limbladian operator. Work on the
study of symmetries of Limbladian operators by one of the authors (F.I.) and
J. Venkatraman is in progress.

The study of the symmetry $su(2)$ of the Kerr oscillator presented here can
be extended to two coupled Kerr oscillators $su(2)_{1}\oplus su(2)_{2}$ in
the same way in which it is done in the proton-neutron interacting boson
model (IBM2) in nuclear physics $su(6)_{1}\oplus su(6)_{2}$ \cite{booknuc},
in triatomic molecules $su(4)_{1}\oplus su(4)_{2}$ \cite{bookmol}, and coupled benders $su(3)_{1}\oplus su(3)_{2}$ \cite{Larese2014}  in molecular physics, and, most importantly, to a large number of coupled oscillators $\sum_{i}\oplus su(2)_{i}$ on a lattice, in the same way in which it is done in the algebraic theory of crystal vibrations \cite{Iachello2015,dietz2017algebraic}, for example in an Ising lattice \cite{goto2019quantum,puri2017quantum,kanao2021high,dykman2018interaction,darmawan_practical_2021}, thus playing an important role in the development
of quantum computers based on Kerr parametric oscillators \cite{goto2019quantum, puri2017}, which is the ultimate goal of the research initiated in this article. The study of the symmetries of two squeeze-driven Kerr
oscillators requires a generalization of the methods of \cite{booknuc, bookmol} to their non-compact versions, in particular, for the
one-dimensional squeeze-driven Kerr oscillator to the coupled algebras $sp(2,\mathbb{R})_{1}\oplus sp(2,\mathbb{R})_{2}$. This study, however, is relatively
straightforward, since the algebra of $sp(2,\mathbb{R})$ is isomorphic to $su(1,1)$,
the non-compact version of the familiar angular momentum algebra. Generalization to nonlinear parametric oscillators with quartic terms $\hat a ^{\dagger 4}\hat a ^4$ or higher nonlinearities are also possible by expanding the symmetry from $su(2)$ to $su(4)$ or $su(n).$

Further applications of the group-theoretic methods and techniques discussed in this paper are to the study of squeeze-driven systems other than the Kerr oscillator, and to their associated quantum phase transitions (QPT) and excited state quantum phase transitions (ESQPT), for example to the squeeze-driven Rabi and Dicke models~\cite{Shen2022,Yang2023} and the Jaynes-Cummings model~\cite{Shen2022}. The algebraic structure of these models is that of $sp(2,\mathbb{R})$ generated by $\hat{a}^{\dagger} \hat{a}, \hat{a}^{\dagger2}, \hat{a}^2$ coupled to $su(2)$ generated by $\sigma_x, \sigma_y, \sigma_z$, that is $sp(2,\mathbb{R}) \oplus su(2)$.

\begin{acknowledgments}
This research was supported by the
NSF CCI grant (Award Number 2124511). 
RGC acknowledges discussions with J. Venkatraman, X. Xiao, M. Devoret, and P. and V. Kurilovich. FPB thanks funding received from the European Union's Horizon 2020 research and innovation program under the Marie
  Sk{\l}odowska-Curie grant agreement No 872081 and from grant
  PID2019-104002GB-C21 funded by MCIN/AEI/ 10.13039/501100011033 and,
  as appropriate, by ``ERDF A way of making Europe'', by the ``European
  Union'' or by the ``European Union NextGenerationEU/PRTR''.
Computing resources supporting
  this work were partially provided by the CEAFMC and Universidad de Huelva High
  Performance Computer (HPC@UHU) located in the Campus Universitario
  el Carmen and funded by FEDER/MINECO project UNHU-15CE-2848.
  \end{acknowledgments}


\begin{thebibliography}{61}%
\makeatletter
\providecommand \@ifxundefined [1]{%
 \@ifx{#1\undefined}
}%
\providecommand \@ifnum [1]{%
 \ifnum #1\expandafter \@firstoftwo
 \else \expandafter \@secondoftwo
 \fi
}%
\providecommand \@ifx [1]{%
 \ifx #1\expandafter \@firstoftwo
 \else \expandafter \@secondoftwo
 \fi
}%
\providecommand \natexlab [1]{#1}%
\providecommand \enquote  [1]{``#1''}%
\providecommand \bibnamefont  [1]{#1}%
\providecommand \bibfnamefont [1]{#1}%
\providecommand \citenamefont [1]{#1}%
\providecommand \href@noop [0]{\@secondoftwo}%
\providecommand \href [0]{\begingroup \@sanitize@url \@href}%
\providecommand \@href[1]{\@@startlink{#1}\@@href}%
\providecommand \@@href[1]{\endgroup#1\@@endlink}%
\providecommand \@sanitize@url [0]{\catcode `\\12\catcode `\$12\catcode
  `\&12\catcode `\#12\catcode `\^12\catcode `\_12\catcode `\%12\relax}%
\providecommand \@@startlink[1]{}%
\providecommand \@@endlink[0]{}%
\providecommand \url  [0]{\begingroup\@sanitize@url \@url }%
\providecommand \@url [1]{\endgroup\@href {#1}{\urlprefix }}%
\providecommand \urlprefix  [0]{URL }%
\providecommand \Eprint [0]{\href }%
\providecommand \doibase [0]{https://doi.org/}%
\providecommand \selectlanguage [0]{\@gobble}%
\providecommand \bibinfo  [0]{\@secondoftwo}%
\providecommand \bibfield  [0]{\@secondoftwo}%
\providecommand \translation [1]{[#1]}%
\providecommand \BibitemOpen [0]{}%
\providecommand \bibitemStop [0]{}%
\providecommand \bibitemNoStop [0]{.\EOS\space}%
\providecommand \EOS [0]{\spacefactor3000\relax}%
\providecommand \BibitemShut  [1]{\csname bibitem#1\endcsname}%
\let\auto@bib@innerbib\@empty
\bibitem [{\citenamefont {Goto}(2016{\natexlab{a}})}]{Goto2016}%
  \BibitemOpen
  \bibfield  {author} {\bibinfo {author} {\bibfnamefont {H.}~\bibnamefont
  {Goto}},\ }\bibfield  {title} {\bibinfo {title} {Bifurcation-based adiabatic
  quantum computation with a nonlinear oscillator network},\ }\href
  {https://doi.org/10.1038/srep21686} {\bibfield  {journal} {\bibinfo
  {journal} {Sci. Rep.}\ }\textbf {\bibinfo {volume} {6}},\ \bibinfo {pages}
  {21686} (\bibinfo {year} {2016}{\natexlab{a}})}\BibitemShut {NoStop}%
\bibitem [{\citenamefont {Goto}(2019)}]{goto2019quantum}%
  \BibitemOpen
  \bibfield  {author} {\bibinfo {author} {\bibfnamefont {H.}~\bibnamefont
  {Goto}},\ }\bibfield  {title} {\bibinfo {title} {Quantum computation based on
  quantum adiabatic bifurcations of {K}err-nonlinear parametric oscillators},\
  }\href@noop {} {\bibfield  {journal} {\bibinfo  {journal} {J. Phys. Soc.
  Japan}\ }\textbf {\bibinfo {volume} {88}},\ \bibinfo {pages} {061015}
  (\bibinfo {year} {2019})}\BibitemShut {NoStop}%
\bibitem [{\citenamefont {Mirrahimi}\ \emph {et~al.}(2014)\citenamefont
  {Mirrahimi}, \citenamefont {Leghtas}, \citenamefont {Albert}, \citenamefont
  {Touzard}, \citenamefont {Schoelkopf}, \citenamefont {Jiang},\ and\
  \citenamefont {Devoret}}]{mirrahimi2014}%
  \BibitemOpen
  \bibfield  {author} {\bibinfo {author} {\bibfnamefont {M.}~\bibnamefont
  {Mirrahimi}}, \bibinfo {author} {\bibfnamefont {Z.}~\bibnamefont {Leghtas}},
  \bibinfo {author} {\bibfnamefont {V.~V.}\ \bibnamefont {Albert}}, \bibinfo
  {author} {\bibfnamefont {S.}~\bibnamefont {Touzard}}, \bibinfo {author}
  {\bibfnamefont {R.~J.}\ \bibnamefont {Schoelkopf}}, \bibinfo {author}
  {\bibfnamefont {L.}~\bibnamefont {Jiang}},\ and\ \bibinfo {author}
  {\bibfnamefont {M.~H.}\ \bibnamefont {Devoret}},\ }\bibfield  {title}
  {\bibinfo {title} {Dynamically protected cat-qubits: a new paradigm for
  universal quantum computation},\ }\href@noop {} {\bibfield  {journal}
  {\bibinfo  {journal} {New J. Phys.}\ }\textbf {\bibinfo {volume} {16}},\
  \bibinfo {pages} {045014} (\bibinfo {year} {2014})}\BibitemShut {NoStop}%
\bibitem [{\citenamefont {Puri}\ \emph
  {et~al.}(2017{\natexlab{a}})\citenamefont {Puri}, \citenamefont {Boutin},\
  and\ \citenamefont {Blais}}]{puri2017}%
  \BibitemOpen
  \bibfield  {author} {\bibinfo {author} {\bibfnamefont {S.}~\bibnamefont
  {Puri}}, \bibinfo {author} {\bibfnamefont {S.}~\bibnamefont {Boutin}},\ and\
  \bibinfo {author} {\bibfnamefont {A.}~\bibnamefont {Blais}},\ }\bibfield
  {title} {\bibinfo {title} {Engineering the quantum states of light in a
  {K}err-nonlinear resonator by two-photon driving},\ }\href@noop {} {\bibfield
   {journal} {\bibinfo  {journal} {npj Quantum Inf.}\ }\textbf {\bibinfo
  {volume} {3}},\ \bibinfo {pages} {1} (\bibinfo {year}
  {2017}{\natexlab{a}})}\BibitemShut {NoStop}%
\bibitem [{\citenamefont {Grimm}\ \emph {et~al.}(2020)\citenamefont {Grimm},
  \citenamefont {Frattini}, \citenamefont {Puri}, \citenamefont {Mundhada},
  \citenamefont {Touzard}, \citenamefont {Mirrahimi}, \citenamefont {Girvin},
  \citenamefont {Shankar},\ and\ \citenamefont {Devoret}}]{grimm2020}%
  \BibitemOpen
  \bibfield  {author} {\bibinfo {author} {\bibfnamefont {A.}~\bibnamefont
  {Grimm}}, \bibinfo {author} {\bibfnamefont {N.~E.}\ \bibnamefont {Frattini}},
  \bibinfo {author} {\bibfnamefont {S.}~\bibnamefont {Puri}}, \bibinfo {author}
  {\bibfnamefont {S.~O.}\ \bibnamefont {Mundhada}}, \bibinfo {author}
  {\bibfnamefont {S.}~\bibnamefont {Touzard}}, \bibinfo {author} {\bibfnamefont
  {M.}~\bibnamefont {Mirrahimi}}, \bibinfo {author} {\bibfnamefont {S.~M.}\
  \bibnamefont {Girvin}}, \bibinfo {author} {\bibfnamefont {S.}~\bibnamefont
  {Shankar}},\ and\ \bibinfo {author} {\bibfnamefont {M.~H.}\ \bibnamefont
  {Devoret}},\ }\bibfield  {title} {\bibinfo {title} {Stabilization and
  operation of a {K}err-cat qubit},\ }\href@noop {} {\bibfield  {journal}
  {\bibinfo  {journal} {Nature}\ }\textbf {\bibinfo {volume} {584}},\ \bibinfo
  {pages} {205} (\bibinfo {year} {2020})}\BibitemShut {NoStop}%
\bibitem [{\citenamefont {Blais}\ \emph {et~al.}(2021)\citenamefont {Blais},
  \citenamefont {Grimsmo}, \citenamefont {Girvin},\ and\ \citenamefont
  {Wallraff}}]{Blais2021}%
  \BibitemOpen
  \bibfield  {author} {\bibinfo {author} {\bibfnamefont {A.}~\bibnamefont
  {Blais}}, \bibinfo {author} {\bibfnamefont {A.~L.}\ \bibnamefont {Grimsmo}},
  \bibinfo {author} {\bibfnamefont {S.~M.}\ \bibnamefont {Girvin}},\ and\
  \bibinfo {author} {\bibfnamefont {A.}~\bibnamefont {Wallraff}},\ }\bibfield
  {title} {\bibinfo {title} {Circuit quantum electrodynamics},\ }\href
  {https://doi.org/10.1103/RevModPhys.93.025005} {\bibfield  {journal}
  {\bibinfo  {journal} {Rev. Mod. Phys.}\ }\textbf {\bibinfo {volume} {93}},\
  \bibinfo {pages} {025005} (\bibinfo {year} {2021})}\BibitemShut {NoStop}%
\bibitem [{\citenamefont {Darmawan}\ \emph {et~al.}(2021)\citenamefont
  {Darmawan}, \citenamefont {Brown}, \citenamefont {Grimsmo}, \citenamefont
  {Tuckett},\ and\ \citenamefont {Puri}}]{darmawan_practical_2021}%
  \BibitemOpen
  \bibfield  {author} {\bibinfo {author} {\bibfnamefont {A.~S.}\ \bibnamefont
  {Darmawan}}, \bibinfo {author} {\bibfnamefont {B.~J.}\ \bibnamefont {Brown}},
  \bibinfo {author} {\bibfnamefont {A.~L.}\ \bibnamefont {Grimsmo}}, \bibinfo
  {author} {\bibfnamefont {D.~K.}\ \bibnamefont {Tuckett}},\ and\ \bibinfo
  {author} {\bibfnamefont {S.}~\bibnamefont {Puri}},\ }\bibfield  {title}
  {\bibinfo {title} {Practical {Quantum} {Error} {Correction} with the {XZZX}
  {Code} and {Kerr}-{Cat} {Qubits}},\ }\href
  {https://doi.org/10.1103/PRXQuantum.2.030345} {\bibfield  {journal} {\bibinfo
   {journal} {PRX Quantum}\ }\textbf {\bibinfo {volume} {2}},\ \bibinfo {pages}
  {030345} (\bibinfo {year} {2021})}\BibitemShut {NoStop}%
\bibitem [{\citenamefont {Kwon}\ \emph {et~al.}(2022)\citenamefont {Kwon},
  \citenamefont {Watabe},\ and\ \citenamefont {Tsai}}]{Kwon2022}%
  \BibitemOpen
  \bibfield  {author} {\bibinfo {author} {\bibfnamefont {S.}~\bibnamefont
  {Kwon}}, \bibinfo {author} {\bibfnamefont {S.}~\bibnamefont {Watabe}},\ and\
  \bibinfo {author} {\bibfnamefont {J.-S.}\ \bibnamefont {Tsai}},\ }\bibfield
  {title} {\bibinfo {title} {Autonomous quantum error correction in a
  four-photon {K}err parametric oscillator},\ }\href
  {https://doi.org/10.1038/s41534-022-00553-z} {\bibfield  {journal} {\bibinfo
  {journal} {npj Quantum Inf.}\ }\textbf {\bibinfo {volume} {8}},\ \bibinfo
  {pages} {40} (\bibinfo {year} {2022})}\BibitemShut {NoStop}%
\bibitem [{\citenamefont {Frattini}\ \emph {et~al.}(2022)\citenamefont
  {Frattini}, \citenamefont {Cortiñas}, \citenamefont {Venkatraman},
  \citenamefont {Xiao}, \citenamefont {Su}, \citenamefont {Lei}, \citenamefont
  {Chapman}, \citenamefont {Joshi}, \citenamefont {Girvin}, \citenamefont
  {Schoelkopf}, \citenamefont {Puri},\ and\ \citenamefont
  {Devoret}}]{Frattini2022}%
  \BibitemOpen
  \bibfield  {author} {\bibinfo {author} {\bibfnamefont {N.~E.}\ \bibnamefont
  {Frattini}}, \bibinfo {author} {\bibfnamefont {R.~G.}\ \bibnamefont
  {Cortiñas}}, \bibinfo {author} {\bibfnamefont {J.}~\bibnamefont
  {Venkatraman}}, \bibinfo {author} {\bibfnamefont {X.}~\bibnamefont {Xiao}},
  \bibinfo {author} {\bibfnamefont {Q.}~\bibnamefont {Su}}, \bibinfo {author}
  {\bibfnamefont {C.~U.}\ \bibnamefont {Lei}}, \bibinfo {author} {\bibfnamefont
  {B.~J.}\ \bibnamefont {Chapman}}, \bibinfo {author} {\bibfnamefont {V.~R.}\
  \bibnamefont {Joshi}}, \bibinfo {author} {\bibfnamefont {S.~M.}\ \bibnamefont
  {Girvin}}, \bibinfo {author} {\bibfnamefont {R.~J.}\ \bibnamefont
  {Schoelkopf}}, \bibinfo {author} {\bibfnamefont {S.}~\bibnamefont {Puri}},\
  and\ \bibinfo {author} {\bibfnamefont {M.~H.}\ \bibnamefont {Devoret}},\
  }\href {https://doi.org/10.48550/ARXIV.2209.03934} {\bibinfo {title} {The
  squeezed {K}err oscillator: spectral kissing and phase-flip robustness}}
  (\bibinfo {year} {2022}),\ \bibinfo {note} {arXiv:2209.03934}\BibitemShut
  {NoStop}%
\bibitem [{\citenamefont {Venkatraman}\ \emph
  {et~al.}(2022{\natexlab{a}})\citenamefont {Venkatraman}, \citenamefont
  {Cortinas}, \citenamefont {Frattini}, \citenamefont {Xiao},\ and\
  \citenamefont {Devoret}}]{Venkatraman2022_Exp}%
  \BibitemOpen
  \bibfield  {author} {\bibinfo {author} {\bibfnamefont {J.}~\bibnamefont
  {Venkatraman}}, \bibinfo {author} {\bibfnamefont {R.~G.}\ \bibnamefont
  {Cortinas}}, \bibinfo {author} {\bibfnamefont {N.~E.}\ \bibnamefont
  {Frattini}}, \bibinfo {author} {\bibfnamefont {X.}~\bibnamefont {Xiao}},\
  and\ \bibinfo {author} {\bibfnamefont {M.~H.}\ \bibnamefont {Devoret}},\
  }\href {https://doi.org/10.48550/ARXIV.2211.04605} {\bibinfo {title} {Quantum
  interference of tunneling paths under a double-well barrier}} (\bibinfo
  {year} {2022}{\natexlab{a}}),\ \Eprint
  {https://arxiv.org/abs/arXiv:2211.04605} {arXiv:2211.04605} \BibitemShut
  {NoStop}%
\bibitem [{\citenamefont {Venkatraman}(2023)}]{VenkatramanThesis}%
  \BibitemOpen
  \bibfield  {author} {\bibinfo {author} {\bibfnamefont {J.}~\bibnamefont
  {Venkatraman}},\ }\emph {\bibinfo {title} {Controlling the Effective
  {H}amiltonian of a Driven Quantum Superconducting Circuit}},\ \href
  {https://www.proquest.com/openview/fd652ab459a37d707aa0fc9f15905927/1?pq-origsite=gscholar&cbl=18750&diss=y}
  {Ph.D. thesis},\ \bibinfo  {school} {Yale University} (\bibinfo {year}
  {2023})\BibitemShut {NoStop}%
\bibitem [{\citenamefont {Kirchmair}\ \emph {et~al.}(2013)\citenamefont
  {Kirchmair}, \citenamefont {Vlastakis}, \citenamefont {Leghtas},
  \citenamefont {Nigg}, \citenamefont {Paik}, \citenamefont {Ginossar},
  \citenamefont {Mirrahimi}, \citenamefont {Frunzio}, \citenamefont {Girvin},\
  and\ \citenamefont {Schoelkopf}}]{Kirchmair2013}%
  \BibitemOpen
  \bibfield  {author} {\bibinfo {author} {\bibfnamefont {G.}~\bibnamefont
  {Kirchmair}}, \bibinfo {author} {\bibfnamefont {B.}~\bibnamefont
  {Vlastakis}}, \bibinfo {author} {\bibfnamefont {Z.}~\bibnamefont {Leghtas}},
  \bibinfo {author} {\bibfnamefont {S.~E.}\ \bibnamefont {Nigg}}, \bibinfo
  {author} {\bibfnamefont {H.}~\bibnamefont {Paik}}, \bibinfo {author}
  {\bibfnamefont {E.}~\bibnamefont {Ginossar}}, \bibinfo {author}
  {\bibfnamefont {M.}~\bibnamefont {Mirrahimi}}, \bibinfo {author}
  {\bibfnamefont {L.}~\bibnamefont {Frunzio}}, \bibinfo {author} {\bibfnamefont
  {S.~M.}\ \bibnamefont {Girvin}},\ and\ \bibinfo {author} {\bibfnamefont
  {R.~J.}\ \bibnamefont {Schoelkopf}},\ }\bibfield  {title} {\bibinfo {title}
  {Observation of quantum state collapse and revival due to the single-photon
  {K}err effect},\ }\href {https://doi.org/10.1038/nature11902} {\bibfield
  {journal} {\bibinfo  {journal} {Nature}\ }\textbf {\bibinfo {volume} {495}},\
  \bibinfo {pages} {205} (\bibinfo {year} {2013})}\BibitemShut {NoStop}%
\bibitem [{\citenamefont {Reh\'ak}\ \emph {et~al.}(2014)\citenamefont
  {Reh\'ak}, \citenamefont {Neilinger}, \citenamefont {Grajcar}, \citenamefont
  {Oelsner}, \citenamefont {H\"ubner}, \citenamefont {Il'ichev},\ and\
  \citenamefont {Meyer}}]{Rehak2014}%
  \BibitemOpen
  \bibfield  {author} {\bibinfo {author} {\bibfnamefont {M.}~\bibnamefont
  {Reh\'ak}}, \bibinfo {author} {\bibfnamefont {P.}~\bibnamefont {Neilinger}},
  \bibinfo {author} {\bibfnamefont {M.}~\bibnamefont {Grajcar}}, \bibinfo
  {author} {\bibfnamefont {G.}~\bibnamefont {Oelsner}}, \bibinfo {author}
  {\bibfnamefont {U.}~\bibnamefont {H\"ubner}}, \bibinfo {author}
  {\bibfnamefont {E.}~\bibnamefont {Il'ichev}},\ and\ \bibinfo {author}
  {\bibfnamefont {H.-G.}\ \bibnamefont {Meyer}},\ }\bibfield  {title} {\bibinfo
  {title} {{Parametric amplification by coupled flux qubits}},\ }\href
  {https://doi.org/10.1063/1.4873719} {\bibfield  {journal} {\bibinfo
  {journal} {Appl. Phys. Lett.}\ }\textbf {\bibinfo {volume} {104}},\ \bibinfo
  {pages} {162604} (\bibinfo {year} {2014})}\BibitemShut {NoStop}%
\bibitem [{\citenamefont {Puri}\ \emph
  {et~al.}(2017{\natexlab{b}})\citenamefont {Puri}, \citenamefont {Andersen},
  \citenamefont {Grimsmo},\ and\ \citenamefont {Blais}}]{puri2017quantum}%
  \BibitemOpen
  \bibfield  {author} {\bibinfo {author} {\bibfnamefont {S.}~\bibnamefont
  {Puri}}, \bibinfo {author} {\bibfnamefont {C.~K.}\ \bibnamefont {Andersen}},
  \bibinfo {author} {\bibfnamefont {A.~L.}\ \bibnamefont {Grimsmo}},\ and\
  \bibinfo {author} {\bibfnamefont {A.}~\bibnamefont {Blais}},\ }\bibfield
  {title} {\bibinfo {title} {Quantum annealing with all-to-all connected
  nonlinear oscillators},\ }\href@noop {} {\bibfield  {journal} {\bibinfo
  {journal} {Nat. commun.}\ }\textbf {\bibinfo {volume} {8}},\ \bibinfo {pages}
  {15785} (\bibinfo {year} {2017}{\natexlab{b}})}\BibitemShut {NoStop}%
\bibitem [{\citenamefont {Goto}\ \emph {et~al.}(2018)\citenamefont {Goto},
  \citenamefont {Lin},\ and\ \citenamefont {Nakamura}}]{Goto2018}%
  \BibitemOpen
  \bibfield  {author} {\bibinfo {author} {\bibfnamefont {H.}~\bibnamefont
  {Goto}}, \bibinfo {author} {\bibfnamefont {Z.}~\bibnamefont {Lin}},\ and\
  \bibinfo {author} {\bibfnamefont {Y.}~\bibnamefont {Nakamura}},\ }\bibfield
  {title} {\bibinfo {title} {Boltzmann sampling from the {I}sing model using
  quantum heating of coupled nonlinear oscillators},\ }\href
  {https://doi.org/10.1038/s41598-018-25492-8} {\bibfield  {journal} {\bibinfo
  {journal} {Sci. Rep.}\ }\textbf {\bibinfo {volume} {8}},\ \bibinfo {pages}
  {7154} (\bibinfo {year} {2018})}\BibitemShut {NoStop}%
\bibitem [{\citenamefont {Amin}\ \emph {et~al.}(2018)\citenamefont {Amin},
  \citenamefont {Andriyash}, \citenamefont {Rolfe}, \citenamefont
  {Kulchytskyy},\ and\ \citenamefont {Melko}}]{Amin2018}%
  \BibitemOpen
  \bibfield  {author} {\bibinfo {author} {\bibfnamefont {M.~H.}\ \bibnamefont
  {Amin}}, \bibinfo {author} {\bibfnamefont {E.}~\bibnamefont {Andriyash}},
  \bibinfo {author} {\bibfnamefont {J.}~\bibnamefont {Rolfe}}, \bibinfo
  {author} {\bibfnamefont {B.}~\bibnamefont {Kulchytskyy}},\ and\ \bibinfo
  {author} {\bibfnamefont {R.}~\bibnamefont {Melko}},\ }\bibfield  {title}
  {\bibinfo {title} {Quantum {B}oltzmann machine},\ }\href
  {https://doi.org/10.1103/PhysRevX.8.021050} {\bibfield  {journal} {\bibinfo
  {journal} {Phys. Rev. X}\ }\textbf {\bibinfo {volume} {8}},\ \bibinfo {pages}
  {021050} (\bibinfo {year} {2018})}\BibitemShut {NoStop}%
\bibitem [{\citenamefont {Yamamoto}\ \emph {et~al.}(2008)\citenamefont
  {Yamamoto}, \citenamefont {Inomata}, \citenamefont {Watanabe}, \citenamefont
  {Matsuba}, \citenamefont {Miyazaki}, \citenamefont {Oliver}, \citenamefont
  {Nakamura},\ and\ \citenamefont {Tsai}}]{Yamamoto2008}%
  \BibitemOpen
  \bibfield  {author} {\bibinfo {author} {\bibfnamefont {T.}~\bibnamefont
  {Yamamoto}}, \bibinfo {author} {\bibfnamefont {K.}~\bibnamefont {Inomata}},
  \bibinfo {author} {\bibfnamefont {M.}~\bibnamefont {Watanabe}}, \bibinfo
  {author} {\bibfnamefont {K.}~\bibnamefont {Matsuba}}, \bibinfo {author}
  {\bibfnamefont {T.}~\bibnamefont {Miyazaki}}, \bibinfo {author}
  {\bibfnamefont {W.~D.}\ \bibnamefont {Oliver}}, \bibinfo {author}
  {\bibfnamefont {Y.}~\bibnamefont {Nakamura}},\ and\ \bibinfo {author}
  {\bibfnamefont {J.~S.}\ \bibnamefont {Tsai}},\ }\bibfield  {title} {\bibinfo
  {title} {{Flux-driven {J}osephson parametric amplifier}},\ }\href
  {https://doi.org/10.1063/1.2964182} {\bibfield  {journal} {\bibinfo
  {journal} {Appl. Phys. Lett.}\ }\textbf {\bibinfo {volume} {93}},\ \bibinfo
  {pages} {042510} (\bibinfo {year} {2008})}\BibitemShut {NoStop}%
\bibitem [{\citenamefont {Bourassa}\ \emph {et~al.}(2012)\citenamefont
  {Bourassa}, \citenamefont {Beaudoin}, \citenamefont {Gambetta},\ and\
  \citenamefont {Blais}}]{Bourassa2012}%
  \BibitemOpen
  \bibfield  {author} {\bibinfo {author} {\bibfnamefont {J.}~\bibnamefont
  {Bourassa}}, \bibinfo {author} {\bibfnamefont {F.}~\bibnamefont {Beaudoin}},
  \bibinfo {author} {\bibfnamefont {J.~M.}\ \bibnamefont {Gambetta}},\ and\
  \bibinfo {author} {\bibfnamefont {A.}~\bibnamefont {Blais}},\ }\bibfield
  {title} {\bibinfo {title} {Josephson-junction-embedded transmission-line
  resonators: From {K}err medium to in-line transmon},\ }\href
  {https://doi.org/10.1103/PhysRevA.86.013814} {\bibfield  {journal} {\bibinfo
  {journal} {Phys. Rev. A}\ }\textbf {\bibinfo {volume} {86}},\ \bibinfo
  {pages} {013814} (\bibinfo {year} {2012})}\BibitemShut {NoStop}%
\bibitem [{\citenamefont {Wustmann}\ and\ \citenamefont
  {Shumeiko}(2013)}]{Wustmann2013}%
  \BibitemOpen
  \bibfield  {author} {\bibinfo {author} {\bibfnamefont {W.}~\bibnamefont
  {Wustmann}}\ and\ \bibinfo {author} {\bibfnamefont {V.}~\bibnamefont
  {Shumeiko}},\ }\bibfield  {title} {\bibinfo {title} {Parametric resonance in
  tunable superconducting cavities},\ }\href
  {https://doi.org/10.1103/PhysRevB.87.184501} {\bibfield  {journal} {\bibinfo
  {journal} {Phys. Rev. B}\ }\textbf {\bibinfo {volume} {87}},\ \bibinfo
  {pages} {184501} (\bibinfo {year} {2013})}\BibitemShut {NoStop}%
\bibitem [{\citenamefont {Krantz}\ \emph {et~al.}(2013)\citenamefont {Krantz},
  \citenamefont {Reshitnyk}, \citenamefont {Wustmann}, \citenamefont
  {Bylander}, \citenamefont {Gustavsson}, \citenamefont {Oliver}, \citenamefont
  {Duty}, \citenamefont {Shumeiko},\ and\ \citenamefont
  {Delsing}}]{Krantz2013}%
  \BibitemOpen
  \bibfield  {author} {\bibinfo {author} {\bibfnamefont {P.}~\bibnamefont
  {Krantz}}, \bibinfo {author} {\bibfnamefont {Y.}~\bibnamefont {Reshitnyk}},
  \bibinfo {author} {\bibfnamefont {W.}~\bibnamefont {Wustmann}}, \bibinfo
  {author} {\bibfnamefont {J.}~\bibnamefont {Bylander}}, \bibinfo {author}
  {\bibfnamefont {S.}~\bibnamefont {Gustavsson}}, \bibinfo {author}
  {\bibfnamefont {W.~D.}\ \bibnamefont {Oliver}}, \bibinfo {author}
  {\bibfnamefont {T.}~\bibnamefont {Duty}}, \bibinfo {author} {\bibfnamefont
  {V.}~\bibnamefont {Shumeiko}},\ and\ \bibinfo {author} {\bibfnamefont
  {P.}~\bibnamefont {Delsing}},\ }\bibfield  {title} {\bibinfo {title}
  {Investigation of nonlinear effects in {J}osephson parametric oscillators
  used in circuit quantum electrodynamics},\ }\href
  {https://doi.org/10.1088/1367-2630/15/10/105002} {\bibfield  {journal}
  {\bibinfo  {journal} {New J. Phys.}\ }\textbf {\bibinfo {volume} {15}},\
  \bibinfo {pages} {105002} (\bibinfo {year} {2013})}\BibitemShut {NoStop}%
\bibitem [{\citenamefont {Eichler}\ and\ \citenamefont
  {Wallraff}(2014)}]{Eichler2014}%
  \BibitemOpen
  \bibfield  {author} {\bibinfo {author} {\bibfnamefont {C.}~\bibnamefont
  {Eichler}}\ and\ \bibinfo {author} {\bibfnamefont {A.}~\bibnamefont
  {Wallraff}},\ }\bibfield  {title} {\bibinfo {title} {Controlling the dynamic
  range of a {J}osephson parametric amplifier},\ }\href
  {https://doi.org/10.1140/epjqt2} {\bibfield  {journal} {\bibinfo  {journal}
  {EPJ Quantum Technology}\ }\textbf {\bibinfo {volume} {1}},\ \bibinfo {pages}
  {2} (\bibinfo {year} {2014})}\BibitemShut {NoStop}%
\bibitem [{\citenamefont {Lin}\ \emph {et~al.}(2014)\citenamefont {Lin},
  \citenamefont {Inomata}, \citenamefont {Koshino}, \citenamefont {Oliver},
  \citenamefont {Nakamura}, \citenamefont {Tsai},\ and\ \citenamefont
  {Yamamoto}}]{Lin2014}%
  \BibitemOpen
  \bibfield  {author} {\bibinfo {author} {\bibfnamefont {Z.~R.}\ \bibnamefont
  {Lin}}, \bibinfo {author} {\bibfnamefont {K.}~\bibnamefont {Inomata}},
  \bibinfo {author} {\bibfnamefont {K.}~\bibnamefont {Koshino}}, \bibinfo
  {author} {\bibfnamefont {W.~D.}\ \bibnamefont {Oliver}}, \bibinfo {author}
  {\bibfnamefont {Y.}~\bibnamefont {Nakamura}}, \bibinfo {author}
  {\bibfnamefont {J.~S.}\ \bibnamefont {Tsai}},\ and\ \bibinfo {author}
  {\bibfnamefont {T.}~\bibnamefont {Yamamoto}},\ }\bibfield  {title} {\bibinfo
  {title} {Josephson parametric phase-locked oscillator and its application to
  dispersive readout of superconducting qubits},\ }\href
  {https://doi.org/10.1038/ncomms5480} {\bibfield  {journal} {\bibinfo
  {journal} {Nat. Comm.}\ }\textbf {\bibinfo {volume} {5}},\ \bibinfo {pages}
  {4480} (\bibinfo {year} {2014})}\BibitemShut {NoStop}%
\bibitem [{\citenamefont {Krantz}\ \emph {et~al.}(2016)\citenamefont {Krantz},
  \citenamefont {Bengtsson}, \citenamefont {Simoen}, \citenamefont
  {Gustavsson}, \citenamefont {Shumeiko}, \citenamefont {Oliver}, \citenamefont
  {Wilson}, \citenamefont {Delsing},\ and\ \citenamefont
  {Bylander}}]{Krantz2016}%
  \BibitemOpen
  \bibfield  {author} {\bibinfo {author} {\bibfnamefont {P.}~\bibnamefont
  {Krantz}}, \bibinfo {author} {\bibfnamefont {A.}~\bibnamefont {Bengtsson}},
  \bibinfo {author} {\bibfnamefont {M.}~\bibnamefont {Simoen}}, \bibinfo
  {author} {\bibfnamefont {S.}~\bibnamefont {Gustavsson}}, \bibinfo {author}
  {\bibfnamefont {V.}~\bibnamefont {Shumeiko}}, \bibinfo {author}
  {\bibfnamefont {W.~D.}\ \bibnamefont {Oliver}}, \bibinfo {author}
  {\bibfnamefont {C.~M.}\ \bibnamefont {Wilson}}, \bibinfo {author}
  {\bibfnamefont {P.}~\bibnamefont {Delsing}},\ and\ \bibinfo {author}
  {\bibfnamefont {J.}~\bibnamefont {Bylander}},\ }\bibfield  {title} {\bibinfo
  {title} {Single-shot read-out of a superconducting qubit using a {J}osephson
  parametric oscillator},\ }\href {https://doi.org/10.1038/ncomms11417}
  {\bibfield  {journal} {\bibinfo  {journal} {Nat. Comm.}\ }\textbf {\bibinfo
  {volume} {7}},\ \bibinfo {pages} {11417} (\bibinfo {year}
  {2016})}\BibitemShut {NoStop}%
\bibitem [{\citenamefont {Dykman}\ and\ \citenamefont
  {Smelyanski}(1990)}]{Dykman1990}%
  \BibitemOpen
  \bibfield  {author} {\bibinfo {author} {\bibfnamefont {M.~I.}\ \bibnamefont
  {Dykman}}\ and\ \bibinfo {author} {\bibfnamefont {V.~N.}\ \bibnamefont
  {Smelyanski}},\ }\bibfield  {title} {\bibinfo {title} {Fluctuational
  transitions between stable states of a nonlinear oscillator driven by random
  resonant force},\ }\href {https://doi.org/10.1103/PhysRevA.41.3090}
  {\bibfield  {journal} {\bibinfo  {journal} {Phys. Rev. A}\ }\textbf {\bibinfo
  {volume} {41}},\ \bibinfo {pages} {3090} (\bibinfo {year}
  {1990})}\BibitemShut {NoStop}%
\bibitem [{\citenamefont {Marthaler}\ and\ \citenamefont
  {Dykman}(2007)}]{Marthaler2007}%
  \BibitemOpen
  \bibfield  {author} {\bibinfo {author} {\bibfnamefont {M.}~\bibnamefont
  {Marthaler}}\ and\ \bibinfo {author} {\bibfnamefont {M.~I.}\ \bibnamefont
  {Dykman}},\ }\bibfield  {title} {\bibinfo {title} {Quantum interference in
  the classically forbidden region: A parametric oscillator},\ }\href
  {https://doi.org/10.1103/PhysRevA.76.010102} {\bibfield  {journal} {\bibinfo
  {journal} {Phys. Rev. A}\ }\textbf {\bibinfo {volume} {76}},\ \bibinfo
  {pages} {010102} (\bibinfo {year} {2007})}\BibitemShut {NoStop}%
\bibitem [{\citenamefont {Dykman}(2012)}]{DykmanBook2012}%
  \BibitemOpen
  \bibfield  {author} {\bibinfo {author} {\bibfnamefont {M.}~\bibnamefont
  {Dykman}},\ }\href
  {https://doi.org/10.1093/acprof:oso/9780199691388.001.0001} {\emph {\bibinfo
  {title} {{F}luctuating {N}onlinear {O}scillators {F}rom {N}anomechanics to
  {Q}uantum {S}uperconducting {C}ircuits}}}\ (\bibinfo  {publisher} {Oxford
  University Press},\ \bibinfo {year} {2012})\BibitemShut {NoStop}%
\bibitem [{\citenamefont {Peano}\ \emph {et~al.}(2012)\citenamefont {Peano},
  \citenamefont {Marthaler},\ and\ \citenamefont {Dykman}}]{Peano2012}%
  \BibitemOpen
  \bibfield  {author} {\bibinfo {author} {\bibfnamefont {V.}~\bibnamefont
  {Peano}}, \bibinfo {author} {\bibfnamefont {M.}~\bibnamefont {Marthaler}},\
  and\ \bibinfo {author} {\bibfnamefont {M.~I.}\ \bibnamefont {Dykman}},\
  }\bibfield  {title} {\bibinfo {title} {Sharp tunneling peaks in a parametric
  oscillator: Quantum resonances missing in the rotating wave approximation},\
  }\href {https://doi.org/10.1103/PhysRevLett.109.090401} {\bibfield  {journal}
  {\bibinfo  {journal} {Phys. Rev. Lett.}\ }\textbf {\bibinfo {volume} {109}},\
  \bibinfo {pages} {090401} (\bibinfo {year} {2012})}\BibitemShut {NoStop}%
\bibitem [{\citenamefont {Lin}\ \emph {et~al.}(2015)\citenamefont {Lin},
  \citenamefont {Nakamura},\ and\ \citenamefont {Dykman}}]{Lin2015}%
  \BibitemOpen
  \bibfield  {author} {\bibinfo {author} {\bibfnamefont {Z.~R.}\ \bibnamefont
  {Lin}}, \bibinfo {author} {\bibfnamefont {Y.}~\bibnamefont {Nakamura}},\ and\
  \bibinfo {author} {\bibfnamefont {M.~I.}\ \bibnamefont {Dykman}},\ }\bibfield
   {title} {\bibinfo {title} {Critical fluctuations and the rates of interstate
  switching near the excitation threshold of a quantum parametric oscillator},\
  }\href {https://doi.org/10.1103/PhysRevE.92.022105} {\bibfield  {journal}
  {\bibinfo  {journal} {Phys. Rev. E}\ }\textbf {\bibinfo {volume} {92}},\
  \bibinfo {pages} {022105} (\bibinfo {year} {2015})}\BibitemShut {NoStop}%
\bibitem [{\citenamefont {Goto}(2016{\natexlab{b}})}]{Goto2016b}%
  \BibitemOpen
  \bibfield  {author} {\bibinfo {author} {\bibfnamefont {H.}~\bibnamefont
  {Goto}},\ }\bibfield  {title} {\bibinfo {title} {Universal quantum
  computation with a nonlinear oscillator network},\ }\href
  {https://doi.org/10.1103/PhysRevA.93.050301} {\bibfield  {journal} {\bibinfo
  {journal} {Phys. Rev. A}\ }\textbf {\bibinfo {volume} {93}},\ \bibinfo
  {pages} {050301} (\bibinfo {year} {2016}{\natexlab{b}})}\BibitemShut
  {NoStop}%
\bibitem [{\citenamefont {Zhang}\ and\ \citenamefont
  {Dykman}(2017)}]{Zhang2017}%
  \BibitemOpen
  \bibfield  {author} {\bibinfo {author} {\bibfnamefont {Y.}~\bibnamefont
  {Zhang}}\ and\ \bibinfo {author} {\bibfnamefont {M.~I.}\ \bibnamefont
  {Dykman}},\ }\bibfield  {title} {\bibinfo {title} {Preparing quasienergy
  states on demand: A parametric oscillator},\ }\href
  {https://doi.org/10.1103/PhysRevA.95.053841} {\bibfield  {journal} {\bibinfo
  {journal} {Phys. Rev. A}\ }\textbf {\bibinfo {volume} {95}},\ \bibinfo
  {pages} {053841} (\bibinfo {year} {2017})}\BibitemShut {NoStop}%
\bibitem [{\citenamefont {Dykman}\ \emph {et~al.}(2018)\citenamefont {Dykman},
  \citenamefont {Bruder}, \citenamefont {L{\"o}rch},\ and\ \citenamefont
  {Zhang}}]{dykman2018interaction}%
  \BibitemOpen
  \bibfield  {author} {\bibinfo {author} {\bibfnamefont {M.~I.}\ \bibnamefont
  {Dykman}}, \bibinfo {author} {\bibfnamefont {C.}~\bibnamefont {Bruder}},
  \bibinfo {author} {\bibfnamefont {N.}~\bibnamefont {L{\"o}rch}},\ and\
  \bibinfo {author} {\bibfnamefont {Y.}~\bibnamefont {Zhang}},\ }\bibfield
  {title} {\bibinfo {title} {Interaction-induced time-symmetry breaking in
  driven quantum oscillators},\ }\href@noop {} {\bibfield  {journal} {\bibinfo
  {journal} {Phys. Rev. B}\ }\textbf {\bibinfo {volume} {98}},\ \bibinfo
  {pages} {195444} (\bibinfo {year} {2018})}\BibitemShut {NoStop}%
\bibitem [{\citenamefont {Roberts}\ and\ \citenamefont
  {Clerk}(2020)}]{Roberts2020}%
  \BibitemOpen
  \bibfield  {author} {\bibinfo {author} {\bibfnamefont {D.}~\bibnamefont
  {Roberts}}\ and\ \bibinfo {author} {\bibfnamefont {A.~A.}\ \bibnamefont
  {Clerk}},\ }\bibfield  {title} {\bibinfo {title} {Driven-dissipative quantum
  {K}err resonators: New exact solutions, photon blockade and quantum
  bistability},\ }\href {https://doi.org/10.1103/PhysRevX.10.021022} {\bibfield
   {journal} {\bibinfo  {journal} {Phys. Rev. X}\ }\textbf {\bibinfo {volume}
  {10}},\ \bibinfo {pages} {021022} (\bibinfo {year} {2020})}\BibitemShut
  {NoStop}%
\bibitem [{\citenamefont {Venkatraman}\ \emph
  {et~al.}(2022{\natexlab{b}})\citenamefont {Venkatraman}, \citenamefont
  {Xiao}, \citenamefont {Corti\~nas}, \citenamefont {Eickbusch},\ and\
  \citenamefont {Devoret}}]{Venkatraman2022_PRL}%
  \BibitemOpen
  \bibfield  {author} {\bibinfo {author} {\bibfnamefont {J.}~\bibnamefont
  {Venkatraman}}, \bibinfo {author} {\bibfnamefont {X.}~\bibnamefont {Xiao}},
  \bibinfo {author} {\bibfnamefont {R.~G.}\ \bibnamefont {Corti\~nas}},
  \bibinfo {author} {\bibfnamefont {A.}~\bibnamefont {Eickbusch}},\ and\
  \bibinfo {author} {\bibfnamefont {M.~H.}\ \bibnamefont {Devoret}},\
  }\bibfield  {title} {\bibinfo {title} {Static effective {H}amiltonian of a
  rapidly driven nonlinear system},\ }\href
  {https://doi.org/10.1103/PhysRevLett.129.100601} {\bibfield  {journal}
  {\bibinfo  {journal} {Phys. Rev. Lett.}\ }\textbf {\bibinfo {volume} {129}},\
  \bibinfo {pages} {100601} (\bibinfo {year} {2022}{\natexlab{b}})}\BibitemShut
  {NoStop}%
\bibitem [{\citenamefont {Xiao}\ \emph {et~al.}(2023)\citenamefont {Xiao},
  \citenamefont {Venkatraman}, \citenamefont {Cortiñas}, \citenamefont
  {Chowdhury},\ and\ \citenamefont {Devoret}}]{xiao2023diagrammatic}%
  \BibitemOpen
  \bibfield  {author} {\bibinfo {author} {\bibfnamefont {X.}~\bibnamefont
  {Xiao}}, \bibinfo {author} {\bibfnamefont {J.}~\bibnamefont {Venkatraman}},
  \bibinfo {author} {\bibfnamefont {R.~G.}\ \bibnamefont {Cortiñas}}, \bibinfo
  {author} {\bibfnamefont {S.}~\bibnamefont {Chowdhury}},\ and\ \bibinfo
  {author} {\bibfnamefont {M.~H.}\ \bibnamefont {Devoret}},\ }\href@noop {}
  {\bibinfo {title} {A diagrammatic method to compute the effective
  {H}amiltonian of driven nonlinear oscillators}} (\bibinfo {year} {2023}),\
  \Eprint {https://arxiv.org/abs/2304.13656} {arXiv:2304.13656} \BibitemShut
  {NoStop}%
\bibitem [{\citenamefont {García-Mata}\ \emph {et~al.}(2023)\citenamefont
  {García-Mata}, \citenamefont {Cortiñas}, \citenamefont {Xiao},
  \citenamefont {Chávez-Carlos}, \citenamefont {Batista}, \citenamefont
  {Santos},\ and\ \citenamefont {Wisniacki}}]{nacho}%
  \BibitemOpen
  \bibfield  {author} {\bibinfo {author} {\bibfnamefont {I.}~\bibnamefont
  {García-Mata}}, \bibinfo {author} {\bibfnamefont {R.~G.}\ \bibnamefont
  {Cortiñas}}, \bibinfo {author} {\bibfnamefont {X.}~\bibnamefont {Xiao}},
  \bibinfo {author} {\bibfnamefont {J.}~\bibnamefont {Chávez-Carlos}},
  \bibinfo {author} {\bibfnamefont {V.~S.}\ \bibnamefont {Batista}}, \bibinfo
  {author} {\bibfnamefont {L.~F.}\ \bibnamefont {Santos}},\ and\ \bibinfo
  {author} {\bibfnamefont {D.~A.}\ \bibnamefont {Wisniacki}},\ }\href@noop {}
  {\bibinfo {title} {Effective versus {F}loquet theory for the {K}err
  parametric oscillator}} (\bibinfo {year} {2023}),\ \Eprint
  {https://arxiv.org/abs/2309.12516} {arXiv:2309.12516} \BibitemShut {NoStop}%
\bibitem [{\citenamefont {Reynoso}\ \emph {et~al.}(2023)\citenamefont
  {Reynoso}, \citenamefont {Nader}, \citenamefont {Ch\'avez-Carlos},
  \citenamefont {Ordaz-Mendoza}, \citenamefont {Corti\~nas}, \citenamefont
  {Batista}, \citenamefont {Lerma-Hern\'andez}, \citenamefont
  {P\'erez-Bernal},\ and\ \citenamefont {Santos}}]{Prado2023}%
  \BibitemOpen
  \bibfield  {author} {\bibinfo {author} {\bibfnamefont {M.~A.~P.}\
  \bibnamefont {Reynoso}}, \bibinfo {author} {\bibfnamefont {D.~J.}\
  \bibnamefont {Nader}}, \bibinfo {author} {\bibfnamefont {J.}~\bibnamefont
  {Ch\'avez-Carlos}}, \bibinfo {author} {\bibfnamefont {B.~E.}\ \bibnamefont
  {Ordaz-Mendoza}}, \bibinfo {author} {\bibfnamefont {R.~G.}\ \bibnamefont
  {Corti\~nas}}, \bibinfo {author} {\bibfnamefont {V.~S.}\ \bibnamefont
  {Batista}}, \bibinfo {author} {\bibfnamefont {S.}~\bibnamefont
  {Lerma-Hern\'andez}}, \bibinfo {author} {\bibfnamefont {F.}~\bibnamefont
  {P\'erez-Bernal}},\ and\ \bibinfo {author} {\bibfnamefont {L.~F.}\
  \bibnamefont {Santos}},\ }\bibfield  {title} {\bibinfo {title} {Quantum
  tunneling and level crossings in the squeeze-driven {K}err oscillator},\
  }\href {https://doi.org/10.1103/PhysRevA.108.033709} {\bibfield  {journal}
  {\bibinfo  {journal} {Phys. Rev. A}\ }\textbf {\bibinfo {volume} {108}},\
  \bibinfo {pages} {033709} (\bibinfo {year} {2023})}\BibitemShut {NoStop}%
\bibitem [{\citenamefont {Chávez-Carlos}\ \emph {et~al.}(2023)\citenamefont
  {Chávez-Carlos}, \citenamefont {Lezama}, \citenamefont {Cortiñas},
  \citenamefont {Venkatraman}, \citenamefont {Devoret}, \citenamefont
  {Batista}, \citenamefont {Pérez-Bernal},\ and\ \citenamefont
  {Santos}}]{Chavez2022}%
  \BibitemOpen
  \bibfield  {author} {\bibinfo {author} {\bibfnamefont {J.}~\bibnamefont
  {Chávez-Carlos}}, \bibinfo {author} {\bibfnamefont {T.~L.~M.}\ \bibnamefont
  {Lezama}}, \bibinfo {author} {\bibfnamefont {R.~G.}\ \bibnamefont
  {Cortiñas}}, \bibinfo {author} {\bibfnamefont {J.}~\bibnamefont
  {Venkatraman}}, \bibinfo {author} {\bibfnamefont {M.~H.}\ \bibnamefont
  {Devoret}}, \bibinfo {author} {\bibfnamefont {V.~S.}\ \bibnamefont
  {Batista}}, \bibinfo {author} {\bibfnamefont {F.}~\bibnamefont
  {Pérez-Bernal}},\ and\ \bibinfo {author} {\bibfnamefont {L.~F.}\
  \bibnamefont {Santos}},\ }\bibfield  {title} {\bibinfo {title} {Spectral
  kissing and its dynamical consequences in the squeeze-driven {K}err
  oscillator},\ }\href {https://doi.org/10.1038/s41534-023-00745-1} {\bibfield
  {journal} {\bibinfo  {journal} {npj Quantum Inf.}\ }\textbf {\bibinfo
  {volume} {9}},\ \bibinfo {pages} {76} (\bibinfo {year} {2023})}\BibitemShut
  {NoStop}%
\bibitem [{\citenamefont {Shen}\ \emph {et~al.}(2022)\citenamefont {Shen},
  \citenamefont {Tang}, \citenamefont {Shi}, \citenamefont {Wu}, \citenamefont
  {Yang},\ and\ \citenamefont {Zheng}}]{Shen2022}%
  \BibitemOpen
  \bibfield  {author} {\bibinfo {author} {\bibfnamefont {L.-T.}\ \bibnamefont
  {Shen}}, \bibinfo {author} {\bibfnamefont {C.-Q.}\ \bibnamefont {Tang}},
  \bibinfo {author} {\bibfnamefont {Z.}~\bibnamefont {Shi}}, \bibinfo {author}
  {\bibfnamefont {H.}~\bibnamefont {Wu}}, \bibinfo {author} {\bibfnamefont
  {Z.-B.}\ \bibnamefont {Yang}},\ and\ \bibinfo {author} {\bibfnamefont
  {S.-B.}\ \bibnamefont {Zheng}},\ }\bibfield  {title} {\bibinfo {title}
  {Squeezed-light-induced quantum phase transition in the jaynes-cummings
  model},\ }\href {https://doi.org/10.1103/PhysRevA.106.023705} {\bibfield
  {journal} {\bibinfo  {journal} {Phys. Rev. A}\ }\textbf {\bibinfo {volume}
  {106}},\ \bibinfo {pages} {023705} (\bibinfo {year} {2022})}\BibitemShut
  {NoStop}%
\bibitem [{\citenamefont {Yang}\ \emph {et~al.}(2023)\citenamefont {Yang},
  \citenamefont {Shi}, \citenamefont {Yang}, \citenamefont {tuo Shen},\ and\
  \citenamefont {Zheng}}]{Yang2023}%
  \BibitemOpen
  \bibfield  {author} {\bibinfo {author} {\bibfnamefont {J.}~\bibnamefont
  {Yang}}, \bibinfo {author} {\bibfnamefont {Z.}~\bibnamefont {Shi}}, \bibinfo
  {author} {\bibfnamefont {Z.-B.}\ \bibnamefont {Yang}}, \bibinfo {author}
  {\bibfnamefont {L.}~\bibnamefont {tuo Shen}},\ and\ \bibinfo {author}
  {\bibfnamefont {S.-B.}\ \bibnamefont {Zheng}},\ }\bibfield  {title} {\bibinfo
  {title} {First-order quantum phase transition in the squeezed {R}abi model},\
  }\href {https://doi.org/10.1088/1402-4896/acc1b4} {\bibfield  {journal}
  {\bibinfo  {journal} {Phys. Scr.}\ }\textbf {\bibinfo {volume} {98}},\
  \bibinfo {pages} {045107} (\bibinfo {year} {2023})}\BibitemShut {NoStop}%
\bibitem [{\citenamefont {Ruiz}\ \emph {et~al.}(2023)\citenamefont {Ruiz},
  \citenamefont {Gautier}, \citenamefont {Guillaud},\ and\ \citenamefont
  {Mirrahimi}}]{ruiz2022}%
  \BibitemOpen
  \bibfield  {author} {\bibinfo {author} {\bibfnamefont {D.}~\bibnamefont
  {Ruiz}}, \bibinfo {author} {\bibfnamefont {R.}~\bibnamefont {Gautier}},
  \bibinfo {author} {\bibfnamefont {J.}~\bibnamefont {Guillaud}},\ and\
  \bibinfo {author} {\bibfnamefont {M.}~\bibnamefont {Mirrahimi}},\ }\bibfield
  {title} {\bibinfo {title} {Two-photon driven {K}err quantum oscillator with
  multiple spectral degeneracies},\ }\href
  {https://doi.org/10.1103/PhysRevA.107.042407} {\bibfield  {journal} {\bibinfo
   {journal} {Phys. Rev. A}\ }\textbf {\bibinfo {volume} {107}},\ \bibinfo
  {pages} {042407} (\bibinfo {year} {2023})}\BibitemShut {NoStop}%
\bibitem [{\citenamefont {Iachello}(1994)}]{Iachello1994}%
  \BibitemOpen
  \bibfield  {author} {\bibinfo {author} {\bibfnamefont {F.}~\bibnamefont
  {Iachello}},\ }\bibfield  {title} {\bibinfo {title} {Lie algebras,
  cohomologies and new applications of quantum mechanics},\ }in\ \href@noop {}
  {\emph {\bibinfo {booktitle} {Contemporary Mathematics}}},\ Vol.\ \bibinfo
  {volume} {160}\ (\bibinfo  {publisher} {American Mathematical Society,
  Providence, RI},\ \bibinfo {year} {1994})\ pp.\ \bibinfo {pages}
  {151--171}\BibitemShut {NoStop}%
\bibitem [{\citenamefont {Iachello}\ and\ \citenamefont
  {Arima}(1987)}]{booknuc}%
  \BibitemOpen
  \bibfield  {author} {\bibinfo {author} {\bibfnamefont {F.}~\bibnamefont
  {Iachello}}\ and\ \bibinfo {author} {\bibfnamefont {A.}~\bibnamefont
  {Arima}},\ }\href@noop {} {\emph {\bibinfo {title} {The Interacting Boson
  Model}}}\ (\bibinfo  {publisher} {Cambridge University Press, Cambridge},\
  \bibinfo {year} {1987})\BibitemShut {NoStop}%
\bibitem [{\citenamefont {Iachello}\ and\ \citenamefont
  {Levine}(1995)}]{bookmol}%
  \BibitemOpen
  \bibfield  {author} {\bibinfo {author} {\bibfnamefont {F.}~\bibnamefont
  {Iachello}}\ and\ \bibinfo {author} {\bibfnamefont {R.~D.}\ \bibnamefont
  {Levine}},\ }\href@noop {} {\emph {\bibinfo {title} {Algebraic Theory of
  Molecules}}}\ (\bibinfo  {publisher} {Oxford University Press, Oxford},\
  \bibinfo {year} {1995})\BibitemShut {NoStop}%
\bibitem [{\citenamefont {Iachello}(2006)}]{bookalg}%
  \BibitemOpen
  \bibfield  {author} {\bibinfo {author} {\bibfnamefont {F.}~\bibnamefont
  {Iachello}},\ }\href@noop {} {\emph {\bibinfo {title} {Lie Algebras and
  Applications (Lecture Notes in Physics)}}},\ Vol.\ \bibinfo {volume} {708}\
  (\bibinfo  {publisher} {Springer, Berlin},\ \bibinfo {year}
  {2006})\BibitemShut {NoStop}%
\bibitem [{\citenamefont {Braak}(2011)}]{Braak2011}%
  \BibitemOpen
  \bibfield  {author} {\bibinfo {author} {\bibfnamefont {D.}~\bibnamefont
  {Braak}},\ }\bibfield  {title} {\bibinfo {title} {Integrability of the {R}abi
  model},\ }\href {https://doi.org/10.1103/PhysRevLett.107.100401} {\bibfield
  {journal} {\bibinfo  {journal} {Phys. Rev. Lett.}\ }\textbf {\bibinfo
  {volume} {107}},\ \bibinfo {pages} {100401} (\bibinfo {year}
  {2011})}\BibitemShut {NoStop}%
\bibitem [{\citenamefont {Albert}\ and\ \citenamefont
  {Jiang}(2014)}]{Albert2014}%
  \BibitemOpen
  \bibfield  {author} {\bibinfo {author} {\bibfnamefont {V.~V.}\ \bibnamefont
  {Albert}}\ and\ \bibinfo {author} {\bibfnamefont {L.}~\bibnamefont {Jiang}},\
  }\bibfield  {title} {\bibinfo {title} {Symmetries and conserved quantities in
  {L}indblad master equations},\ }\href
  {https://doi.org/10.1103/PhysRevA.89.022118} {\bibfield  {journal} {\bibinfo
  {journal} {Phys. Rev. A}\ }\textbf {\bibinfo {volume} {89}},\ \bibinfo
  {pages} {022118} (\bibinfo {year} {2014})}\BibitemShut {NoStop}%
\bibitem [{\citenamefont {Caprio}\ \emph {et~al.}(2008)\citenamefont {Caprio},
  \citenamefont {Cejnar},\ and\ \citenamefont {Iachello}}]{Caprio2008}%
  \BibitemOpen
  \bibfield  {author} {\bibinfo {author} {\bibfnamefont {M.~S.}\ \bibnamefont
  {Caprio}}, \bibinfo {author} {\bibfnamefont {P.}~\bibnamefont {Cejnar}},\
  and\ \bibinfo {author} {\bibfnamefont {F.}~\bibnamefont {Iachello}},\
  }\bibfield  {title} {\bibinfo {title} {Excited state quantum phase
  transitions in many-body systems},\ }\href
  {https://doi.org/https://doi.org/10.1016/j.aop.2007.06.011} {\bibfield
  {journal} {\bibinfo  {journal} {Ann. Phys.}\ }\textbf {\bibinfo {volume}
  {323}},\ \bibinfo {pages} {1106 } (\bibinfo {year} {2008})}\BibitemShut
  {NoStop}%
\bibitem [{\citenamefont {Cejnar}\ and\ \citenamefont
  {Str\'ansk\'y}(2008)}]{Cejnar2008}%
  \BibitemOpen
  \bibfield  {author} {\bibinfo {author} {\bibfnamefont {P.}~\bibnamefont
  {Cejnar}}\ and\ \bibinfo {author} {\bibfnamefont {P.}~\bibnamefont
  {Str\'ansk\'y}},\ }\bibfield  {title} {\bibinfo {title} {Impact of quantum
  phase transitions on excited-level dynamics},\ }\href
  {https://doi.org/10.1103/PhysRevE.78.031130} {\bibfield  {journal} {\bibinfo
  {journal} {Phys. Rev. E}\ }\textbf {\bibinfo {volume} {78}},\ \bibinfo
  {pages} {031130} (\bibinfo {year} {2008})}\BibitemShut {NoStop}%
\bibitem [{\citenamefont {Cejnar}\ \emph {et~al.}(2021)\citenamefont {Cejnar},
  \citenamefont {Str{\'{a}}nsk{\'{y}}}, \citenamefont {Macek},\ and\
  \citenamefont {Kloc}}]{Cejnar2021}%
  \BibitemOpen
  \bibfield  {author} {\bibinfo {author} {\bibfnamefont {P.}~\bibnamefont
  {Cejnar}}, \bibinfo {author} {\bibfnamefont {P.}~\bibnamefont
  {Str{\'{a}}nsk{\'{y}}}}, \bibinfo {author} {\bibfnamefont {M.}~\bibnamefont
  {Macek}},\ and\ \bibinfo {author} {\bibfnamefont {M.}~\bibnamefont {Kloc}},\
  }\bibfield  {title} {\bibinfo {title} {Excited-state quantum phase
  transitions},\ }\href {https://doi.org/10.1088/1751-8121/abdfe8} {\bibfield
  {journal} {\bibinfo  {journal} {J. Phys. A: Math. Theor.}\ }\textbf {\bibinfo
  {volume} {54}},\ \bibinfo {pages} {133001} (\bibinfo {year}
  {2021})}\BibitemShut {NoStop}%
\bibitem [{\citenamefont {Gilmore}(1974)}]{Gilmore1974}%
  \BibitemOpen
  \bibfield  {author} {\bibinfo {author} {\bibfnamefont {R.}~\bibnamefont
  {Gilmore}},\ }\href@noop {} {\emph {\bibinfo {title} {Lie Groups, {L}ie
  Algebras, and Some of Their Applications}}}\ (\bibinfo  {publisher} {John
  Wiley \& Sons Inc., New York},\ \bibinfo {year} {1974})\BibitemShut {NoStop}%
\bibitem [{\citenamefont {van Roosmalen}(1982)}]{onophd}%
  \BibitemOpen
  \bibfield  {author} {\bibinfo {author} {\bibfnamefont {O.~S.}\ \bibnamefont
  {van Roosmalen}},\ }\emph {\bibinfo {title} {Algebraic Description of Nuclear
  and Molecular Rotation-Vibration Spectra}},\ \href@noop {} {Ph.D. thesis},\
  \bibinfo  {school} {University of Groningen, The Netherlands} (\bibinfo
  {year} {1982})\BibitemShut {NoStop}%
\bibitem [{\citenamefont {Lipkin}\ \emph {et~al.}(1965)\citenamefont {Lipkin},
  \citenamefont {Meshkov},\ and\ \citenamefont {Glick}}]{Lipkin1965}%
  \BibitemOpen
  \bibfield  {author} {\bibinfo {author} {\bibfnamefont {H.~J.}\ \bibnamefont
  {Lipkin}}, \bibinfo {author} {\bibfnamefont {N.}~\bibnamefont {Meshkov}},\
  and\ \bibinfo {author} {\bibfnamefont {A.~J.}\ \bibnamefont {Glick}},\
  }\bibfield  {title} {\bibinfo {title} {Validity of many-body approximation
  methods for a solvable model},\ }\href@noop {} {\bibfield  {journal}
  {\bibinfo  {journal} {Nucl. Phys.}\ }\textbf {\bibinfo {volume} {62}},\
  \bibinfo {pages} {188 } (\bibinfo {year} {1965})}\BibitemShut {NoStop}%
\bibitem [{\citenamefont {Wybourne}(1974)}]{Wybourne1974}%
  \BibitemOpen
  \bibfield  {author} {\bibinfo {author} {\bibfnamefont {B.~G.}\ \bibnamefont
  {Wybourne}},\ }\href@noop {} {\emph {\bibinfo {title} {Classical Groups for
  Physicists}}}\ (\bibinfo  {publisher} {John Wiley \& Sons Inc., New York},\
  \bibinfo {year} {1974})\BibitemShut {NoStop}%
\bibitem [{\citenamefont {Heiss}\ \emph {et~al.}(2005)\citenamefont {Heiss},
  \citenamefont {Scholtz},\ and\ \citenamefont {Geyer}}]{Heiss2005}%
  \BibitemOpen
  \bibfield  {author} {\bibinfo {author} {\bibfnamefont {W.~D.}\ \bibnamefont
  {Heiss}}, \bibinfo {author} {\bibfnamefont {F.~G.}\ \bibnamefont {Scholtz}},\
  and\ \bibinfo {author} {\bibfnamefont {H.~B.}\ \bibnamefont {Geyer}},\
  }\bibfield  {title} {\bibinfo {title} {The large {N} behaviour of the
  {L}ipkin model and exceptional points},\ }\href
  {https://doi.org/10.1088/0305-4470/38/9/002} {\bibfield  {journal} {\bibinfo
  {journal} {J. Phys. A: Math. and General}\ }\textbf {\bibinfo {volume}
  {38}},\ \bibinfo {pages} {1843} (\bibinfo {year} {2005})}\BibitemShut
  {NoStop}%
\bibitem [{\citenamefont {Santos}\ \emph {et~al.}(2016)\citenamefont {Santos},
  \citenamefont {T\'avora},\ and\ \citenamefont {P\'erez-Bernal}}]{Santos2016}%
  \BibitemOpen
  \bibfield  {author} {\bibinfo {author} {\bibfnamefont {L.~F.}\ \bibnamefont
  {Santos}}, \bibinfo {author} {\bibfnamefont {M.}~\bibnamefont {T\'avora}},\
  and\ \bibinfo {author} {\bibfnamefont {F.}~\bibnamefont {P\'erez-Bernal}},\
  }\bibfield  {title} {\bibinfo {title} {Excited-state quantum phase
  transitions in many-body systems with infinite-range interaction:
  Localization, dynamics, and bifurcation},\ }\href
  {https://doi.org/10.1103/PhysRevA.94.012113} {\bibfield  {journal} {\bibinfo
  {journal} {Phys. Rev. A}\ }\textbf {\bibinfo {volume} {94}},\ \bibinfo
  {pages} {012113} (\bibinfo {year} {2016})}\BibitemShut {NoStop}%
\bibitem [{\citenamefont {Nader}\ \emph {et~al.}(2021)\citenamefont {Nader},
  \citenamefont {Gonz\'alez-Rodr\'{\i}guez},\ and\ \citenamefont
  {Lerma-Hern\'andez}}]{Nader2021}%
  \BibitemOpen
  \bibfield  {author} {\bibinfo {author} {\bibfnamefont {D.~J.}\ \bibnamefont
  {Nader}}, \bibinfo {author} {\bibfnamefont {C.~A.}\ \bibnamefont
  {Gonz\'alez-Rodr\'{\i}guez}},\ and\ \bibinfo {author} {\bibfnamefont
  {S.}~\bibnamefont {Lerma-Hern\'andez}},\ }\bibfield  {title} {\bibinfo
  {title} {Avoided crossings and dynamical tunneling close to excited-state
  quantum phase transitions},\ }\href
  {https://doi.org/10.1103/PhysRevE.104.064116} {\bibfield  {journal} {\bibinfo
   {journal} {Phys. Rev. E}\ }\textbf {\bibinfo {volume} {104}},\ \bibinfo
  {pages} {064116} (\bibinfo {year} {2021})}\BibitemShut {NoStop}%
\bibitem [{\citenamefont {Miano}\ \emph {et~al.}(2023)\citenamefont {Miano},
  \citenamefont {Joshi}, \citenamefont {Liu}, \citenamefont {Dai},
  \citenamefont {Parakh}, \citenamefont {Frunzio},\ and\ \citenamefont
  {Devoret}}]{miano2023hamiltonian}%
  \BibitemOpen
  \bibfield  {author} {\bibinfo {author} {\bibfnamefont {A.}~\bibnamefont
  {Miano}}, \bibinfo {author} {\bibfnamefont {V.~R.}\ \bibnamefont {Joshi}},
  \bibinfo {author} {\bibfnamefont {G.}~\bibnamefont {Liu}}, \bibinfo {author}
  {\bibfnamefont {W.}~\bibnamefont {Dai}}, \bibinfo {author} {\bibfnamefont
  {P.~D.}\ \bibnamefont {Parakh}}, \bibinfo {author} {\bibfnamefont
  {L.}~\bibnamefont {Frunzio}},\ and\ \bibinfo {author} {\bibfnamefont {M.~H.}\
  \bibnamefont {Devoret}},\ }\href@noop {} {\bibinfo {title} {Hamiltonian
  extrema of an arbitrary flux-biased {J}osephson circuit}} (\bibinfo {year}
  {2023}),\ \Eprint {https://arxiv.org/abs/2302.03155} {arXiv:2302.03155}
  \BibitemShut {NoStop}%
\bibitem [{\citenamefont {Larese}\ \emph {et~al.}(2014)\citenamefont {Larese},
  \citenamefont {Caprio}, \citenamefont {Pérez-Bernal},\ and\ \citenamefont
  {Iachello}}]{Larese2014}%
  \BibitemOpen
  \bibfield  {author} {\bibinfo {author} {\bibfnamefont {D.}~\bibnamefont
  {Larese}}, \bibinfo {author} {\bibfnamefont {M.~A.}\ \bibnamefont {Caprio}},
  \bibinfo {author} {\bibfnamefont {F.}~\bibnamefont {Pérez-Bernal}},\ and\
  \bibinfo {author} {\bibfnamefont {F.}~\bibnamefont {Iachello}},\ }\bibfield
  {title} {\bibinfo {title} {A study of the bending motion in tetratomic
  molecules by the algebraic operator expansion method},\ }\href
  {https://doi.org/10.1063/1.4856115} {\bibfield  {journal} {\bibinfo
  {journal} {J. Chem. Phys.}\ }\textbf {\bibinfo {volume} {140}},\ \bibinfo
  {pages} {014304} (\bibinfo {year} {2014})}\BibitemShut {NoStop}%
\bibitem [{\citenamefont {Iachello}\ \emph {et~al.}(2015)\citenamefont
  {Iachello}, \citenamefont {Dietz}, \citenamefont {Miski-Oglu},\ and\
  \citenamefont {Richter}}]{Iachello2015}%
  \BibitemOpen
  \bibfield  {author} {\bibinfo {author} {\bibfnamefont {F.}~\bibnamefont
  {Iachello}}, \bibinfo {author} {\bibfnamefont {B.}~\bibnamefont {Dietz}},
  \bibinfo {author} {\bibfnamefont {M.}~\bibnamefont {Miski-Oglu}},\ and\
  \bibinfo {author} {\bibfnamefont {A.}~\bibnamefont {Richter}},\ }\bibfield
  {title} {\bibinfo {title} {Algebraic theory of crystal vibrations:
  Singularities and zeros in vibrations of one- and two-dimensional lattices},\
  }\href {https://doi.org/10.1103/PhysRevB.91.214307} {\bibfield  {journal}
  {\bibinfo  {journal} {Phys. Rev. B}\ }\textbf {\bibinfo {volume} {91}},\
  \bibinfo {pages} {214307} (\bibinfo {year} {2015})}\BibitemShut {NoStop}%
\bibitem [{\citenamefont {Dietz}\ \emph {et~al.}(2017)\citenamefont {Dietz},
  \citenamefont {Iachello},\ and\ \citenamefont {Macek}}]{dietz2017algebraic}%
  \BibitemOpen
  \bibfield  {author} {\bibinfo {author} {\bibfnamefont {B.}~\bibnamefont
  {Dietz}}, \bibinfo {author} {\bibfnamefont {F.}~\bibnamefont {Iachello}},\
  and\ \bibinfo {author} {\bibfnamefont {M.}~\bibnamefont {Macek}},\ }\bibfield
   {title} {\bibinfo {title} {Algebraic theory of crystal vibrations:
  Localization properties of wave functions in two-dimensional lattices},\
  }\href@noop {} {\bibfield  {journal} {\bibinfo  {journal} {Crystals}\
  }\textbf {\bibinfo {volume} {7}},\ \bibinfo {pages} {246} (\bibinfo {year}
  {2017})}\BibitemShut {NoStop}%
\bibitem [{\citenamefont {Kanao}\ and\ \citenamefont
  {Goto}(2021)}]{kanao2021high}%
  \BibitemOpen
  \bibfield  {author} {\bibinfo {author} {\bibfnamefont {T.}~\bibnamefont
  {Kanao}}\ and\ \bibinfo {author} {\bibfnamefont {H.}~\bibnamefont {Goto}},\
  }\bibfield  {title} {\bibinfo {title} {High-accuracy {I}sing machine using
  {K}err-nonlinear parametric oscillators with local four-body interactions},\
  }\href@noop {} {\bibfield  {journal} {\bibinfo  {journal} {npj Quantum Inf.}\
  }\textbf {\bibinfo {volume} {7}},\ \bibinfo {pages} {18} (\bibinfo {year}
  {2021})}\BibitemShut {NoStop}%
\end{thebibliography}
%

\end{document}